\documentclass[aps,prb,reprint,superscriptaddress]{revtex4-2}
\usepackage{blindtext}
\usepackage{siunitx}
\usepackage{graphicx}
\usepackage{natbib}
\usepackage{amsmath}
\usepackage{amssymb}
\usepackage{ bbold }
\usepackage[utf8]{inputenc}
\usepackage{braket}
\usepackage{color}
\usepackage[table]{xcolor}
\usepackage{placeins}
\RequirePackage[T1]{fontenc}
\RequirePackage[utf8]{inputenc}
\usepackage{tikz-cd}
\usepackage{xypic}

\usepackage[colorlinks=true]{hyperref}
\usepackage[toc,page]{appendix}

\newcommand{\rfig}[1]{Fig.\,\ref{#1}}
\newcommand{\rapp}[1]{App.\,\ref{#1}}
\newcommand{\req}[1]{Eq.\,(\ref{#1})}

\newcommand{\rtab}[1]{Tab.\,\ref{#1}}

\begin{document}
\title{Error channels in quantum nondemolition measurements on spin systems}
\author{Benjamin Joecker}
\affiliation{Centre for Quantum Computation and Communication Technology, School of Electrical Engineering \& Telecommunications, UNSW Sydney, NSW, 2052, Australia}
\author{Holly G. Stemp}
\affiliation{Centre for Quantum Computation and Communication Technology, School of Electrical Engineering \& Telecommunications, UNSW Sydney, NSW, 2052, Australia}
\author{Irene Fern\'andez de Fuentes}
\affiliation{Centre for Quantum Computation and Communication Technology, School of Electrical Engineering \& Telecommunications, UNSW Sydney, NSW, 2052, Australia}
\author{Mark A. I. Johnson}
\affiliation{Centre for Quantum Computation and Communication Technology, School of Electrical Engineering \& Telecommunications, UNSW Sydney, NSW, 2052, Australia}
\author{Andrea Morello}
\affiliation{Centre for Quantum Computation and Communication Technology, School of Electrical Engineering \& Telecommunications, UNSW Sydney, NSW, 2052, Australia}

\begin{abstract}
Quantum nondemolition (QND) measurements are a precious resource for quantum information processing. Repetitive QND measurements can boost the fidelity of qubit preparation and measurement, even when the underlying single-shot measurements are of low fidelity. However, this fidelity boost is limited by the degree in which the physical system allows for a truly QND process -- slight deviations from ideal QND measurement result in bit flip errors (`quantum jumps') if the measurement is repeated too often. Here, we develop a theoretical framework to understand and quantify the resulting error arising from deviation from perfect QND measurement in model spin qubit systems. We first develop our model on the ubiquitous example of exchange-coupled electron spins qubits tunnel-coupled to a charge reservoir. We then extend it to electron-nuclear spin systems, to illustrate the crucial similarities and differences between the two limits. Applied to the well-understood platform of a donor nuclear spin in silicon, the model shows excellent agreement with experiments. For added generality, we conclude the work by considering the effect of anisotropic spin couplings. \end{abstract}

\maketitle

\section{Introduction}
The measurement postulate of quantum mechanics is usually described as follows: Upon measuring some physical quantity $\mathcal{A}$, described by a Hermitian operator $\hat{A}$, the outcome of the measurement can only be one of the eigenvalues $a_n$ of $\hat{A}$; immediately after a measurement has occurred, the system will be found in the eigenstate $\ket{\phi_n}$ associated with the eigenvalue $a_n$ \cite{cohen1977quantum}. Despite being found in all textbooks, the second part of the postulate does not describe all (or even the majority of) practical situations. For example, measuring the presence or absence of a photon can destroy the photon completely -- if a photon is registered by the detector, we know that one photon existed before the measurement, but after the measurement it no longer does. A similar scenario is found in the measurement of electron spins in semiconductors, via the mechanism of energy-dependent tunneling \cite{elzerman2004single,xiao2004electrical,morello2010single}. A high-energy electron (spin-up, in materials with positive Land\'{e} $g$-factor) can tunnel into a cold charge reservoir, leaving an imprint in a nearby charge sensor. After the measurement, however, the spin-up electron is entirely lost -- all we are left with is the knowledge that the now-lost electron was in the spin-up state. 

The textbook example, where the system remains in the post-measurement state corresponding to the observed eigenvalue, describes what is otherwise known as a quantum nondemolition (QND) measurement\cite{braginsky1996quantum}. Its key property is that, since the first measurement with outcome $a_n$ projects the system in the eigenstate $\ket{\phi_n}$ of the observable under study, every subsequent measurement will return the same outcome $a_n$ with certainty. In realistic experiments, where there may be noise affecting the apparatus and reducing the single-shot fidelity, the ability to perform repetitive QND measurements and averaging over the results can greatly enhance the overall fidelity of the outcome\cite{wallraff2005approaching,vijay2011observation,neumann2010single,pla2013high,muhonen2014storing}. A high-fidelity QND readout enables equally reliable state preparation\cite{philips2022universal}. Other applications of robust QND measurements include quantum error correction\cite{terhal2015quantum}, entanglement by measurement \cite{ruskov2003entanglement}, or observation of the quantum Zeno effect\cite{itano1990quantum,gambetta2008quantum}. 

The experimental realization of a QND measurement usually consists of three components: (i) the quantum system of interest (here assumed to be a qubit), described by the Hamiltonian $\hat{H}_{\rm Q}$, (ii) an ancillary quantum system, $\hat{H}_{\rm A}$, which can be read out (destructively or otherwise), and (iii) a coupling between the two systems, $\hat{H}_{\rm C}$. The condition a measurement must fulfill to behave in QND measurement manner is that $\hat{H}_{\rm Q}$ commutes with the interaction $\hat{H}_{\rm  C}$\cite{braginsky1996quantum,pla2013high}
\begin{align}
    \left[\hat{H}_{\rm C},\hat{H}_{\rm Q}\right]=0.
    \label{Eq:QNDcondition}
\end{align}

A faithful implementation of this condition can be achieved in the context of cavity quantum electrodynamics (cQED), by coupling the qubit to photons in a high-quality resonator \cite{mabuchi2002cavity,blais2021circuit}. In the dispersive limit, the qubit state only shifts the cavity frequency, which is measurable without affecting the qubit state itself\cite{koch2007charge}. A wide range of QND experiments in solid-state cQED setups have been demonstrated on superconducting \cite{wallraff2005approaching,vijay2011observation} and spin-based\cite{mi2018coherent,samkharadze2018strong} qubits. The cQED architecture has been extremely successful in enabling intermediate-scale quantum computation \cite{arute2019quantum,wu2021strong}, but requires a large footprint due to the size of the resonator.

Here, we study an approach to realize a QND readout, where both the qubit and the ancillary system are spins hosted in a solid-state device. In the context of semiconductor spin qubits, this approach has the benefit of retaining the small footprint and high qubit density that is typical of such platforms\cite{vandersypen2017interfacing}. This kind of QND measurement has been experimentally demonstrated with electrons in quantum dots\cite{nakajima2019quantum,xue2020repetitive,philips2022universal}, and in various electron-nuclear spin systems including Nitrogen-Vacancy (NV) centers in diamond\cite{jiang2009repetitive,neumann2010single,robledo2011high,waldherr2011dark,dreau2013single,lovchinsky2016nuclear}, donors in silicon\cite{pla2013high,muhonen2014storing,pla2014coherent}, and a quantum dot electron spin coupled to a nuclear spin ensemble\cite{dyte2023quantum}.

An ideal QND measurement would require the interaction between the qubit and the ancilla to be of Ising type, or a similar interaction that fulfills \req{Eq:QNDcondition}. Unfortunately, the native coupling mechanism in semiconductor spin systems is either the Heisenberg exchange (between pairs of electron spins) or the hyperfine interaction (between an electron and a nuclear spin). These do not commute with the Hamiltonian of the qubit, and thus violate the QND condition in \req{Eq:QNDcondition}. During the QND protocol, these interactions repeatedly weakly entangle the qubit to the ancilla, which in turn is measured projectively. This process can lead to unwanted flips of the data qubit, and constitutes an error channel for the QND readout.

The goal of this paper is to develop a general theoretical framework to understand the error channels in this type of QND measurements, and quantify the corresponding error rates. 
The paper is organized as follows. In section 2 we review how to perform QND measurements on the idealized  example of two Ising exchange-coupled spins, and establish theoretical models describing the relevant processes. Section 3 discusses error channels in a realistic system of two Heisenberg exchange-coupled spins. Section 4 applies our models to donor spin systems, covering both exchange-coupled donor electrons and a nuclear spin hyperfine-coupled to a bound electron, in the isotropic and the anisotropic case. Section 5 concludes and summarizes the results.

\section{Ising exchange: Ideal QND readout scheme}
\label{Sec:QND}
\begin{figure}
    \includegraphics[width=1.0\columnwidth]{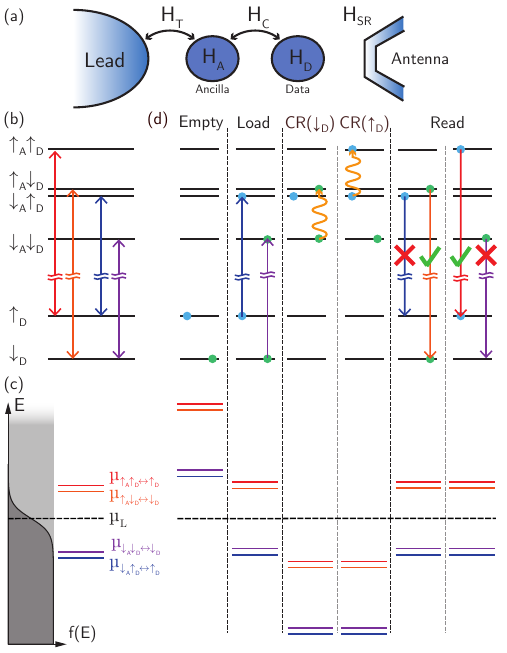}
    \caption{Illustration of the quantum nondemolition (QND) readout scheme of an electron spin qubit. (a) The model under consideration. A data qubit (with Hamiltonian $\hat{H}_{\rm D}$) is coupled to an ancilla ($\hat{H}_{\rm A}$) via a coupling $\hat{H}_{\rm C}$. The ancilla is tunnel-coupled ($\hat{H}_{\rm T}$) to a lead. The state of the ancilla can be flipped by driving a spin resonance (SR) antenna ($\hat{H}_{\rm SR}$). (b) The energy levels of the two-particle (top) and one-particle (bottom) states are split by the ancilla on-site energy and the mutual charging energy with the data qubit. For Ising-coupled spins (\req{Eq:Ising}) $\hat{H}_{\rm T}$ only allows the indicated transitions (\req{Eq:SelectionRule}). (c) Exemplary Fermi distribution of the lead with respect to the chemical potentials from (b). Placing the Fermi energy of the lead between the $\uparrow_{\rm A}$ and $\downarrow_{\rm A}$ transitions allows for initialization (Load) and readout of the ancilla (Read) via spin-dependent tunneling from and to the lead. (d) Going from left to right illustrates the different stages of the QND measurement, where the blue (green) dot indicates the state evolution in case of a $\uparrow_{\rm D}$ ($\downarrow_{\rm D}$). The flip of the ancilla can be conditional on the $\downarrow_{\rm D}$ (CR$(\downarrow_{\rm D})$) or $\uparrow_{\rm D}$ (CR$(\uparrow_{\rm D})$).}
    \label{Fig:1}
\end{figure}

The setup needed to perform the QND measurement scheme is shown in \rfig{Fig:1}a. A data qubit, described by the Hamiltonian $\hat{H}_{\rm D}$, is connected to an ancilla ($\hat{H}_{\rm A}$) via a coupling $\hat{H}_{\rm C}$. The ancilla in turn is tunnel-coupled ($\hat{H}_{\rm T}$) to a lead, i.e. a cold reservoir of electrons, for the purpose of performing spin readout based on energy-dependent tunneling\cite{elzerman2004single,xiao2004electrical,morello2010single}. The same mechanism ensures that, after readout, the ancilla is reset to the low-energy spin state. The ancilla spin can be controlled via spin resonance ($\hat{H}_{\rm SR}$). In the following, we establish the model for the data-ancilla system and treat $\hat{H}_{\rm T}$ as a perturbation to determine tunnel rates from and to the lead.

The data qubit and the ancilla are described by
\begin{align}
    \hat{H}_{\rm D} = \sum_{\sigma} (E_{\rm D}+\epsilon_{\rm D}\sigma)\hat{d}_\sigma^\dagger\hat{d}_\sigma\\
    \hat{H}_{\rm A} = \sum_{\sigma} (E_{\rm A}+\epsilon_{\rm A}\sigma)\hat{a}_\sigma^\dagger\hat{a}_\sigma
\end{align}
where $\hat{d}_\sigma^\dagger$ ($\hat{a}_\sigma^\dagger$) creates a particle with spin $\sigma\in\{\uparrow,\downarrow\}$ on the data (ancilla). $E_{\rm D/A}$ is the on-site energy and $\epsilon_{\rm D/A}$ the corresponding spin splitting. We assume that $\epsilon_{\rm D}$ and $\epsilon_{\rm A}$ are close in value but sufficiently different to ensure that both spins can be individually addressed via spin resonance. This is easily achieved in quantum dot systems by using gradient magnetic fields from micromagnets\cite{zajac2018resonantly} or exploiting different $g$-factors between dots\cite{huang2019fidelity}; in donor systems, it can be achieved by orienting the donors' nuclear spins in opposite directions\cite{kalra2014robust,madzik2021conditional}. Since the data qubit electron is never removed, in the following we may drop the constant energy offset $E_{\rm D}$. Allowing no more than single particle occupation, the Ising coupling between two spins is described by
\begin{align}
    \hat{H}_{\rm C} = \sum_{\sigma,\sigma^\prime} (E_{\rm C}+J_{\rm I}\sigma\sigma^\prime)\hat{n}_{\rm a,\sigma}\hat{n}_{\rm d,\sigma^\prime},
\end{align}
where $\hat{n}_{\rm a,\sigma}=\hat{a}_\sigma^\dagger\hat{a}_\sigma$ is the number operator for an ancilla with spin $\sigma$ and $\hat{n}_{\rm d,\sigma}$ the equivalent for the data. $E_{\rm C}$ is the mutual charging energy and $J_{\rm I}$ the magnitude of the Ising spin-coupling. Throughout this paper we express energies in units of frequency by dividing by the Planck constant $h$. The resulting energy levels are sketched in \rfig{Fig:1}b for $E_{\rm C}+E_{\rm A}\gg\epsilon_{\rm A}\gtrsim\epsilon_{\rm D}\gg J_{\rm I}$. The energy splitting between the center of the two-particle (top) and one-particle (bottom) states is given by $E_{\rm C}+E_{\rm A}$.

From the point of view of the data-ancilla system, the tunneling of a particle to (from) the lead is the annihilation (creation) of an ancilla particle. A full picture including an explicit treatment of the lead can be found in \rapp{App:Lead}. Assuming the tunnel coupling $t_{0}$ is constant over the relevant energy range, i.e. the tunnel barrier is not changing much with energy (E), the rate of a transition between the one-particle (1P) and two-particle (2P) states scales with $|\braket{{\rm 1P}|\sum_{\sigma} t_{0} \hat{a}_{\sigma}|{\rm 2P}}|^2$ (see \rapp{App:Lead}). The 1P states are simply $\ket{\uparrow_{\textrm{D}}}$ and $\ket{\downarrow_{\textrm{D}}}$, while the 2P states can be easily found by solving the effective spin Hamiltonian in the two-particle subspace
\begin{align}
    \hat{H}_{\rm A}+\hat{H}_{\rm D}+\hat{H}_{\rm C}=\epsilon_{\rm A}\hat{S}_{\textrm{A},z}+\epsilon_{\rm D}\hat{S}_{\textrm{D},z}+J_{\rm I}\hat{S}_{\textrm{A},z}\hat{S}_{\textrm{D},z}
    \label{Eq:Ising}
\end{align}
Here, $\hat{S}_{\textrm{A/D},z}$ are the spin operators of the ancilla/data qubit in the basis $\{\ket{\uparrow_{\textrm{A/D}}},\ket{\downarrow_{\textrm{A/D}}}\}$. Since all terms in \req{Eq:Ising} commute, the eigenstates of the data-ancilla system are separable, i.e. there is no entanglement between the two spins. The only allowed transitions are those that preserve the spin of the data qubit (colored arrows in \rfig{Fig:1}b) as required in a perfect QND measurement.

The chemical potentials $\mu_{{\rm 2P}\leftrightarrow {\rm 1P}}$, i.e. the energy change of the system that the ancilla particle carries as it tunnels into the lead, are indicated by the length of the colored arrows in \rfig{Fig:1}b. Energetically, a particle can only tunnel off (on) the ancilla, if a state at this energy is unoccupied (occupied) in the lead. To quantify this, we treat the lead as a continuum of states with density $n(E)$. The fraction of states in the lead occupied at energy $E$ for a given temperature $T$ is given by the Fermi distribution,
\begin{align}
    f(E)=\frac{1}{1+e^{(E-\mu_{\rm L})/k_{\rm B}T}},
    \label{Eq:Fermi}
\end{align}
where $\mu_{\rm L}$ is the chemical potential of the lead and $k_{\rm B}$ Boltzmann's constant (in units of Hz/K, given our choice of units for the energy). The full transition rates at which a particle can tunnel between the lead and the ancilla are then given by the golden rule\cite{cota2003spin,golovach2004transport,johnson2022beating,osika2022shelving}
\begin{subequations}
\begin{align}
    \Gamma^{\rm in}_{\rm 1P\rightarrow 2P}&=|\braket{{\rm 1P}|\sum_{\sigma} t_{0} \hat{a}_{\sigma}|{\rm 2P}}|^2 n(\mu_{{\rm 1P}\leftrightarrow {\rm 2P}})f(\mu_{\rm 1P\leftrightarrow 2P})
    \label{Eq:LoadRate}\\
    \Gamma^{\rm out}_{\rm 2P\rightarrow 1P}&=|\braket{{\rm 2P}|\sum_{\sigma} t_{0} \hat{a}^\dagger_{\sigma}|{\rm 1P}}|^2 n(\mu_{{\rm 1P}\leftrightarrow {\rm 2P}})(1-f(\mu_{\rm 1P\leftrightarrow 2P})).
    \label{Eq:UnloadRate}
\end{align}
\label{Eq:GoldenRates}
\end{subequations}
Here, \req{Eq:LoadRate} describes the tunneling of an electron onto the ancilla while \req{Eq:UnloadRate} is the inverse process. \rfig{Fig:1}c shows an exemplary Fermi distribution with respect to the transition energies to the right. Electrostatically tuning the lead with respect to the ancilla, i.e. tuning $\mu_{\rm 2P\leftrightarrow 1P}-\mu_{\rm L}$ in \req{Eq:Fermi}, shifts the ratio of occupied and empty states. A transition in tune with occupied states (dark grey shade) will result in $\Gamma^{\rm in}_{\rm 1P\rightarrow 2P}\gg\Gamma^{\rm out}_{\rm 2P\rightarrow 1P}$, while a transition in tune with empty states (light grey shade) leads to $\Gamma^{\rm out}_{\rm 2P\rightarrow 1P}\gg\Gamma^{\rm in}_{\rm 1P\rightarrow 2P}$.

Finally, the spin resonance antenna\cite{dehollain2012nanoscale} depicted on the right in \rfig{Fig:1}a provides an oscillating magnetic field $B_1$ that drives spin resonance transitions via the Hamiltonian $\hat{H}_{\rm SR}=\gamma B_1 \cos{(2\pi \nu t)}(\hat{S}_{\textrm{A},x}+\hat{S}_{\textrm{D},x})$, where $\nu$ is a frequency corresponding to one of the conditional rotations (CR) shown in \rfig{Fig:1}d and $\gamma$ the gyromagentic ratio of the spin (in units of Hz/T).

The QND measurement scheme now follows the steps outlined in \rfig{Fig:1}d. At the start, a previous QND measurement has left the data qubit either in the $\uparrow$ or $\downarrow$ state, marked by the blue and green circles. In the first step, we load a $\downarrow$ electron onto the ancilla by tuning the corresponding chemical potential below that of the lead, while keeping the $\uparrow$ chemical potential well above. Subsequently, we tune (`plunge') to a regime where tunnel-off events are suppressed, and perform a $\pi$-pulse on the ancilla conditional on the state of the data qubit. Here, we can choose to either drive a CR dependent on the data qubit being $\downarrow$ (CR$(\downarrow_{\rm D})$) or $\uparrow$ (CR$(\uparrow_{\rm D})$), or alternating between the two (CR$(\updownarrow_{\rm D})$). Finally, we readout the ancilla spin by pulsing back to the load/read configuration, where now only an $\uparrow$ ancilla will cause a tunnel event, detectable by a charge sensor. In the CR$(\downarrow_{\rm D})$ case, a tunnel event confirms that the data qubit was previously in the $\downarrow$ state; in the CR$(\uparrow_{\rm D})$ case, finding a $\uparrow$ ancilla is associated with a $\uparrow$ data qubit. Note that if no tunnel event occurs, we have already performed the load stage for the next QND measurement, while if a tunnel event occurs a $\downarrow$ electron will automatically be reloaded onto the ancilla if we wait for a sufficiently long time at the load/read tuning.

The readout protocol with Ising-coupled spins described above can suffer from various imperfections, such as state preparation and measurement errors on the ancilla due to the nonzero temperature of the lead\cite{johnson2022beating}, or imperfect $\pi$-pulses in the implementation of the conditional rotations. Fortunately, such errors preserve the QND nature of the measurement and leave the data qubit unaffected. Therefore, it is possible to repeat the QND measurement cycle multiple times to increase the confidence in determining the qubit's state\cite{neumann2010single,pla2013high,nakajima2019quantum,xue2020repetitive}.

\section{Heisenberg exchange: Breakdown of the QND condition}
\begin{figure*}
    \includegraphics[width=2.0\columnwidth]{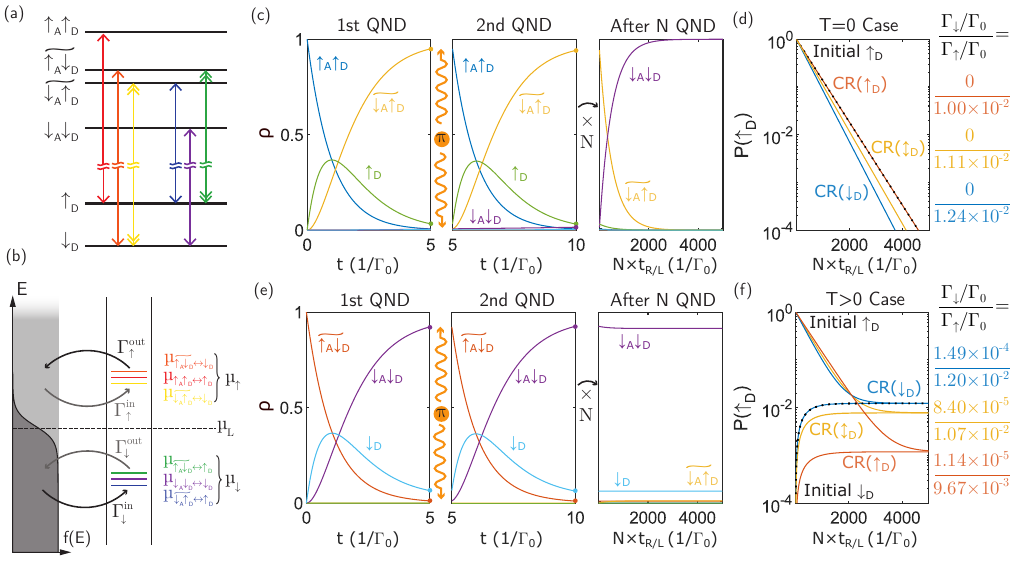}
    \caption{Error channels in the QND measurement for Heisenberg exchange coupled spins. (a) Energy levels of the two-particle (top) and one-particle (bottom) states. For exchange-coupled spins (\req{Eq:Exchange}) the antiparallel spin states are hybridized (\req{Eq:HybridStates}) and $\hat{H}_{\rm T}$ allows the additional transitions indicated by the yellow and green double arrow (\req{Eq:SelectionRule}). (b) Chemical potentials with respect to an exemplary Fermi distribution in the lead for the read/load tuning. The resulting rates in \req{Eq:BaseRates} belonging to an up ($\mu_{\uparrow}$) or down ($\mu_{\downarrow}$) spin particle tunneling in or out are indicated. (c) Simulations of the QND measurement with an initial $\uparrow_{\rm D}$, and driving a conditional rotation on the ancilla dependent on the data being $\uparrow_{\rm D}$ (CR$(\uparrow_{\rm D})$) for $T=\SI{0}{}$ (\req{Eq:Liouvillian2e0K}) with $\Gamma^{\rm in}_{\downarrow}=\Gamma^{\rm out}_{\uparrow}=\Gamma_0$, and $s^2=\SI{2.5e-3}{}$. The first two panels show the state distribution as a function of time during the first two read/load periods. The last panel gives the final distribution after the Nth repetition, i.e. the last point of each QND step (indicated by the small dots in the first two panels) spaced by $t_{\rm R/L}=5/\Gamma_0$. In total $N=1000$ QND cycles are simulated. (d) Probability of finding the data in a $\uparrow_{\rm D}$ state (\req{Eq:Pup}) after repeated QND measurements as a function of time on a logarithmic scale. The cases of driving the CR$(\uparrow_{\rm D})$, CR$(\downarrow_{\rm D})$, and alternating between the two (CR$(\updownarrow_{\rm D})$) are shown. The case of CR$(\uparrow_{\rm D})$ and an initial $\uparrow_{\rm D}$ discussed in (c) is highlighted by the black dotted line. The data qubit flipping rates on the right are (in multiples of the bare tunnel rate $\Gamma_0$) extracted by fitting \req{Eq:Fitting}. The evolution of the data qubit is well described by these rates, independently of its initial state. In the $T=0$ limit only an initial $\uparrow_{\rm D}$ can be flipped. (e) and (f) are the same as (c) and (d), but for $T>0$, i.e. $\Gamma^{\rm in}_{\downarrow}=\Gamma^{\rm out}_{\uparrow}=\Gamma_0(1-f)$ and $\Gamma^{\rm in}_{\uparrow}=\Gamma^{\rm out}_{\downarrow}=\Gamma_0f$ with $f=0.03$. (e) Evolution of an initial $\downarrow_{\rm D}$, using CR$(\downarrow_{\rm D})$. (f) In the $T>0$ case, an initial $\downarrow_{\rm D}$ can be flipped as well.}
    \label{Fig:2}
\end{figure*}

The idealized QND measurement described in the previous section relied upon an Ising-type interaction between the spins. In the near-totality of practical applications, however, the true interaction takes the form of Heisenberg exchange:
\begin{align}
    \hat{H}_{\rm A}+\hat{H}_{\rm D}+\hat{H}_{\rm C}=\epsilon_{\rm A}\hat{S}_{\textrm{A},z}+\epsilon_{\rm D}\hat{S}_{\textrm{D},z}+J_{\rm H}\Vec{S}_{\textrm{A}}\cdot\Vec{S}_{\textrm{D}},
    \label{Eq:Exchange}
\end{align}
where $J_{\rm H}$ is the amplitude of the Heisenberg exchange interaction and $\Vec{S}_{\textrm{A/D}}=(\hat{S}_{\textrm{A/D},x},\hat{S}_{\textrm{A/D},y},\hat{S}_{\textrm{A/D},z})$ the vector of spin operators. The Heisenberg interaction Hamiltonian contains terms that do not commute with the data qubit Hamiltonian, and therefore violates the QND condition \req{Eq:QNDcondition}. 

\rfig{Fig:2}a shows the 1P and 2P energy level in this case for $\epsilon_{\rm A}\gtrsim\epsilon_{\rm D}\gg J_{\rm H}$. Crucially, the exchange coupling now weakly entangles the antiparallel spin states, resulting in eigenstates of the form
\begin{subequations}
\begin{align}
    \ket{\widetilde{\uparrow_{\rm A}\downarrow_{\rm D}}}&=c\ket{\uparrow_{\rm A}\downarrow_{\rm D}}-s\ket{\downarrow_{\rm A}\uparrow_{\rm D}}
    \label{Eq:HybridStateud}
    \\
    \ket{\widetilde{\downarrow_{\rm A}\uparrow_{\rm D}}}&=c\ket{\downarrow_{\rm A}\uparrow_{\rm D}}+s\ket{\uparrow_{\rm A}\downarrow_{\rm D}},
    \label{Eq:HybridStatedu}
\end{align}
\label{Eq:HybridStates}
\end{subequations}
where $c=\cos(\theta)$ and $s=\sin(\theta)$ with $\tan(2\theta)=J_{\rm H}/\Delta\epsilon$ and $\Delta\epsilon=\epsilon_{\rm A}-\epsilon_{\rm D}$. This hybridization enables additional spin transitions marked by the double-headed yellow and green arrows in \rfig{Fig:2}a associated with flipping the data qubit spin, i.e. a QND violation. To analyze the probability with which these flips occur while the ancilla is loaded and read out, we need to find the associated tunnel rates in \req{Eq:GoldenRates}. As discussed in the previous section, these have two main contributions.

First, we have the amplitudes of the transition matrix elements coupling the 1P and 2P states, given by
\begin{align}
    M_{\rm 1P,2P}=|\braket{{\rm 1P}|\hat{a}_{\uparrow}+\hat{a}_{\downarrow}|{\rm 2P}}|^2.
    \label{Eq:SelectionRule}
\end{align}
These can be viewed as a selection rule, as transitions in \rfig{Fig:2}a are only allowed if $M_{\rm 1P,2P}$ is nonzero.

Secondly, we have the selection through energy. At the read/load position shown in \rfig{Fig:2}b, we now have three high-energy transitions associated with an $\uparrow$ particle tunneling, and three low-energy transitions where a $\downarrow$ particle tunnels. Note that, since $\Delta\epsilon\ll\epsilon_{\rm A},\epsilon_{\rm D}$, transitions where the data and ancilla are exchanged in the 2P state have roughly the same energy. For the rest of this section, we will assume that the density of states in the lead is approximately constant over the relevant energy scale and we may define a bare tunneling rate $\Gamma_0=|t_0|^2n(\mu)$ with $n(\mu)=const$. Assuming further that $f(E)$ is approximately the same for all the three $\uparrow$ as well as all three $\downarrow$ transitions, the relevant rates are
\begin{subequations}
\begin{align}
    &\Gamma^{\rm in}_{\uparrow}=\Gamma_0f(\mu_{\uparrow})\\
    &\Gamma^{\rm out}_{\uparrow}=\Gamma_0(1-f(\mu_{\uparrow}))\\
    &\Gamma^{\rm in}_{\downarrow}=\Gamma_0f(\mu_{\downarrow})\\
    &\Gamma^{\rm out}_{\downarrow}=\Gamma_0(1-f(\mu_{\downarrow})),
\end{align}
\label{Eq:BaseRates}
\end{subequations}
where $\mu_{\uparrow}$ ($\mu_{\downarrow}$) represents the high (low) energy chemical potentials. Finally, we also assume that $\mu_{\rm L}$ is perfectly centered between $\mu_{\uparrow}$ and $\mu_{\downarrow}$, so that $f=f(\mu_{\uparrow})=1-f(\mu_{\downarrow})$.

\subsection{Zero-temperature limit}

We begin by analyzing the $T=0$ limit, where $f(E)$ is a step function and only $\Gamma^{\rm in}_{\downarrow}$ and $\Gamma^{\rm out}_{\uparrow}$ are non-zero at the read/load position. To understand the breakdown of the QND measurement, we consider the isolated process of a particle being loaded into the ancilla while the data qubit is $\uparrow_{\rm D}$. The relevant transitions rates are:
\begin{subequations}
\begin{align}
    \Gamma^{\rm in}_{\uparrow_{\rm D}\rightarrow \widetilde{\uparrow_{\rm A}\downarrow_{\rm D}}}&=\Gamma^{\rm in}_{\downarrow} M_{\uparrow_{\rm D},\widetilde{\uparrow_{\rm A}\downarrow_{\rm D}}}=\Gamma^{\rm in}_{\downarrow} s^2,\\
    \Gamma^{\rm in}_{\uparrow_{\rm D}\rightarrow \widetilde{\downarrow_{\rm A}\uparrow_{\rm D}}}&=\Gamma^{\rm in}_{\downarrow} M_{\uparrow_{\rm D},\widetilde{\downarrow_{\rm A}\uparrow_{\rm D}}}=\Gamma^{\rm in}_{\downarrow} c^2.
\end{align}
\label{Eq:TunnelInRates}
\end{subequations}
The time evolution of the system then obeys the rate equations\cite{johnson2022beating,osika2022shelving}
\begin{align}
    \dot{\Vec{\rho}}=\mathcal{L}\Vec{\rho},
\end{align}
where $\Vec{\rho}$ is the vector of the state distribution and $\mathcal{L}$ is the Liouvillian of the system. We first restrict our analysis to the three states in \req{Eq:TunnelInRates} and obtain
\begin{align}
    \begin{pmatrix}
    \dot{\rho}_{\widetilde{\uparrow_{\rm A}\downarrow_{\rm D}}}\\
    \dot{\rho}_{\widetilde{\downarrow_{\rm A}\uparrow_{\rm D}}}\\
    \dot{\rho}_{\uparrow_{\rm D}}
    \end{pmatrix}
    =
    \begin{pmatrix}
    0&0&\Gamma^{\rm in}_{\downarrow}s^2\\
    0&0&\Gamma^{\rm in}_{\downarrow}c^2\\
    0&0&-\Gamma^{\rm in}_{\downarrow}
    \end{pmatrix}
    \begin{pmatrix}
    \rho_{\widetilde{\uparrow_{\rm A}\downarrow_{\rm D}}}\\
    \rho_{\widetilde{\downarrow_{\rm A}\uparrow_{\rm D}}}\\
    \rho_{\uparrow_{\rm D}}
    \end{pmatrix}.
    \label{Eq:RateLoadee}
\end{align}
For an initial $\rho_{\uparrow_{\rm D}}(0)=1$, we find the probabilities of the 2P states as
\begin{subequations}
\begin{align}
    \rho_{\widetilde{\uparrow_{\rm A}\downarrow_{\rm D}}}(t)=&s^2\left(1-e^{-\Gamma^{\rm in}_{\downarrow}t}\right)
    \xrightarrow{t\rightarrow\infty}s^2,\\
    \rho_{\widetilde{\downarrow_{\rm A}\uparrow_{\rm D}}}(t)=&c^2\left(1-e^{-\Gamma^{\rm in}_{\downarrow}t}\right)
    \xrightarrow{t\rightarrow\infty}c^2.
\end{align}
\end{subequations}
After waiting long enough for an electron to tunnel, we have a non-zero probability to flip the data during the load. For a ratio $J_{\rm H}/\Delta\epsilon \approx 1/10$, as typically found in experiments\cite{zajac2018resonantly,huang2019fidelity,madzik2021conditional,joecker2021full}, the probability is
\begin{align}
    s^2=\sin^2\left(\frac{1}{2}\tan^{-1}\left(\frac{J_{\rm H}}{\Delta\epsilon}\right)\right)\approx\left(\frac{J_{\rm H}}{2\Delta\epsilon}\right)^2=\SI{2.5e-3}{}.
    \label{Eq:PJcase}
\end{align}
This represents the probability of accidentally flipping the data qubit, purely as a consequence of it becoming weakly entangled with the ancilla qubit each time it is loaded. Further errors can arise due to the ancilla electron tunneling back out, leaving the subspace of states in \req{Eq:RateLoadee}.

The full QND readout protocol includes repetitive ancilla loading and readout, and rotation of the ancilla conditional on a specific data qubit state. Simulating the full protocol requires solving the rate equations including all states and transitions allowed by \req{Eq:SelectionRule} indicated in \rfig{Fig:2}a. The full Liouvillian in the zero temperature limit is given in \req{Eq:Liouvillian2e0K} in \rapp{App:A}. 
The QND protocol with an initial $\uparrow_{\rm D}$ data qubit and the CR$(\uparrow_D)$ ancilla rotation is described in \rfig{Fig:2}c. After loading the $\widetilde{\downarrow_{\rm A}\uparrow_{\rm D}}$ state and performing the CR$(\uparrow_D)$ operation the system is in the state $\uparrow_{\rm A}\uparrow_{\rm D}$. The first panel shows the evolution of the state distribution as a function of time in the first read/load window. The $\uparrow_{\rm A}\uparrow_{\rm D}$ probability decreases as the ancilla tunnels out with rate $\Gamma^{\rm out}_{\uparrow}=\Gamma_0$, leading to an initially increased probability of finding the 1P state $\uparrow_{\rm D}$. The $\uparrow_{\rm D}$ probability quickly peaks and decreases as the ancilla is reloaded into the $\widetilde{\downarrow_{\rm A}\uparrow_{\rm D}}$ state. Here, we chose a read/load window of duration $t_{\rm R/L}=5/\Gamma_0$ in which a $\uparrow_{\rm A}$ ancilla is read out and replaced by a $\downarrow_{\rm A}$ with a probability of 0.96. Not surprisingly, the shape of $\rho_{\uparrow_{\rm D}}(t)$ reflects the averaged signal from a charge sensor detecting an initial $\uparrow_{\rm A}$ electron (see e.g. the supplementary information of Ref.~\cite{morello2010single} for an example).

The second panel shows the repetition of this process after the application of the conditional $\pi$-pulse that swaps the $\widetilde{\downarrow_{\rm A}\uparrow_{\rm D}}$ and $\uparrow_{\rm A}\uparrow_{\rm D}$ populations. Here, we notice a small nonzero $\rho_{\downarrow_{\rm A}\downarrow_{\rm D}}$ population from the start, since the data qubit may have already accidentally flipped. The last panel shows the final state distribution after the Nth repetition of the QND scheme, where each repetition takes $t=5/\Gamma_0$. In the shown case, the data qubit will eventually flip to $\downarrow_{\rm D}$ and remain in that state.

\rfig{Fig:2}d shows the probability of finding the data in an $\uparrow_{\rm D}$ state independent of the current state of the ancilla
\begin{align}
    P(\uparrow_{\rm D})=\rho_{\uparrow_{\rm A}\uparrow_{\rm D}}+\rho_{\widetilde{\downarrow_{\rm A}\uparrow_{\rm D}}}+\rho_{\uparrow_{\rm D}}
    \label{Eq:Pup}
\end{align}
after the Nth read/load window on a logarithmic scale and for the cases of driving the CR$(\uparrow_D)$, CR$(\downarrow_D)$, and alternating between the two (CR$(\updownarrow_D)$). We can see that an initial $P(\uparrow_{\rm D})=1$ decays exponentially as a function of time. The decay is in fact fastest when the ancilla rotation is conditioned on the opposite state (CR$(\downarrow_D)$) and the system practically idles in the $\widetilde{\downarrow_{\rm A}\uparrow_{\rm D}}$ state, from which the data qubit can be flipped via $\widetilde{\downarrow_{\rm A}\uparrow_{\rm D}}\rightarrow\downarrow_{\rm D}$. In the CR$(\uparrow_D)$ case, shown in \rfig{Fig:2}c, the system spends time in the $\uparrow_{\rm A}\uparrow_{\rm D}$, which only allows spin-preserving transitions, and the decay is slowed down. Fitting \req{Eq:Fitting} to the evolution we extract the flipping rates on the right of \rfig{Fig:2}d. An initial $\downarrow_{\rm D}$ data qubit state cannot be flipped due to the fact that the corresponding transitions require an $\uparrow_{\rm A}$ ($\downarrow_{\rm A}$) particle to tunnel into (out of) the system. These transitions are energetically forbidden at zero temperature.


\subsection{Nonzero temperature}
In the $T>0$ case, the Fermi distribution is no longer a step function, resulting in small but non-zero $\Gamma^{\rm in}_{\uparrow}$ and $\Gamma^{\rm out}_{\downarrow}$ and the Liouvillian in \req{Eq:Liouvillian2e0K} needs to be extended by the elements in \req{Eq:Liouvillian2efK}. To demonstrate how this introduces flips of a $\downarrow_{\rm D}$, \rfig{Fig:2}e shows the same simulations as in \rfig{Fig:2}c for the case of driving the CR$(\downarrow_D)$, an initial $\downarrow_{\rm D}$, and  $T>0$, i.e.  $\Gamma^{\rm in}_{\downarrow}=\Gamma^{\rm out}_{\uparrow}=\Gamma_0(1-f)$ and $\Gamma^{\rm in}_{\uparrow}=\Gamma^{\rm out}_{\downarrow}=\Gamma_0 f$. We choose $f=0.03$, corresponding to $\mu_\uparrow-\mu_\downarrow=2\mu_B\times\SI{1}{T}\approx\SI{27.97}{\giga\hertz}$ (for electron spins with Land\'{e} $g$-factor $\approx 2$) and $k_B T=k_B\times\SI{200}{\milli\kelvin}\approx\SI{4}{\giga\hertz}$. Here, the read/load process is marginally slower compared to the $T=0$ case, due to the slight reduction of the fast rates ($\Gamma^{\rm out}_{\uparrow}$ and $\Gamma^{\rm in}_{\downarrow}$) through temperature and the small probability of accidentally loading a $\uparrow_{\rm A}$ or a loaded $\downarrow_{\rm A}$ tunneling back out. This reduces the probability to reinitialize the system to $\downarrow_{\rm A}\downarrow_{\rm D}$ to 0.92, leaving the ancilla empty with probability 0.07. After repeated measurements, the probability of finding the data flipped to $\uparrow_{\rm D}$, i.e. the $\widetilde{\downarrow_{\rm A}\uparrow_{\rm D}}$ state, increases, but then quickly saturates at $\SI{1.2e-2}{}$. While a $\downarrow_{\rm D}$ can now be flipped via processes involving the slow rates $\Gamma^{\rm in}_{\uparrow}$ and $\Gamma^{\rm out}_{\downarrow}$, the much faster processes involving $\Gamma^{\rm in}_{\downarrow}$ and $\Gamma^{\rm out}_{\uparrow}$ will still flip the $\uparrow_{\rm D}$ back down, and the rate equations converge to an equilibrium state of these two competing processes.

The $P(\uparrow_{\rm D})$ (see \req{Eq:Pup}) as a function of QND repetitions for $T>0$ in \rfig{Fig:2}f show that a different equilibrium is reached depending on whether the CR$(\uparrow_{\rm D})$ or CR$(\downarrow_{\rm D})$ is driven. The equilibrium ratio $P(\uparrow_D)/(1-P(\uparrow_D))$ equals the ratios of the $\uparrow_{\rm D}$ and $\downarrow_{\rm D}$ flipping rates (see \rapp{App:B}). The data qubit flipping rates for $T>0$ on right in \rfig{Fig:2}f are extracted by fitting \req{Eq:Fitting}. The $\Gamma_\downarrow$ flipping rates are much slower when driving the CR$(\downarrow_{\rm D})$ transition. This differences can be understood by looking at the flow chart of the involved processes. In the case of CR$(\downarrow_{\rm D})$, flipping the data from $\downarrow_{\rm D}$ to $\uparrow_{\rm D}$ requires a slow and a fast process:
\begin{align}
\xymatrix{
  &\uparrow_{\rm D}\ar[r]^{\rm fast}& \widetilde{\downarrow_{\rm A}\uparrow_{\rm D}}\\
\widetilde{\uparrow_{\rm A}\downarrow_{\rm D}} \ar[r]_{\rm fast} \ar[ru]^{\rm slow} & \downarrow_{\rm D}\ar[ru]_{\rm slow} & 
} 
\end{align}
Conversely, driving CR$(\uparrow_{\rm D})$ requires two slow processes 
\begin{align}
\xymatrix{
  \downarrow_{\rm A}\downarrow_{\rm D}\ar[r]_{\rm slow}& \downarrow_{\rm D}\ar[r]_{\rm slow}&\widetilde{\downarrow_{\rm A}\uparrow_{\rm D}},
}
\end{align}
which results in a much slower flipping rate. Alternating between the two (CR$(\updownarrow_{\rm D})$), yields an intermediate situation.

Overall, the above discussion indicates that driving the CR$(\uparrow_{\rm D})$ transition yields a better QND readout fidelity, and is particularly effective at protecting the $\downarrow_{\rm D}$ state.

\section{Application to donor spins and electron-nuclear systems}
The previous section assumed the data and ancilla qubits to be exchange-coupled electron spins in a double quantum dot. In this section we apply the discussion to the case where the qubits are electron spins bound to donors atoms in silicon\cite{morello2020donor,madzik2021conditional}, and then extend it to the case where the data qubit is the donor nuclear spin\cite{pla2013high}, while the ancilla is the donor-bound electron\cite{pla2013high}. Finally, we extend the latter scenario to the case where the electron-nuclear hyperfine coupling is anisotropic, which applies to much wider range of electron-nuclear spin systems, including NV-centers in diamond\cite{neumann2010single}.

\subsection{Exchange-coupled donor electron spins}

\begin{figure*}
    \includegraphics[width=2.0\columnwidth]{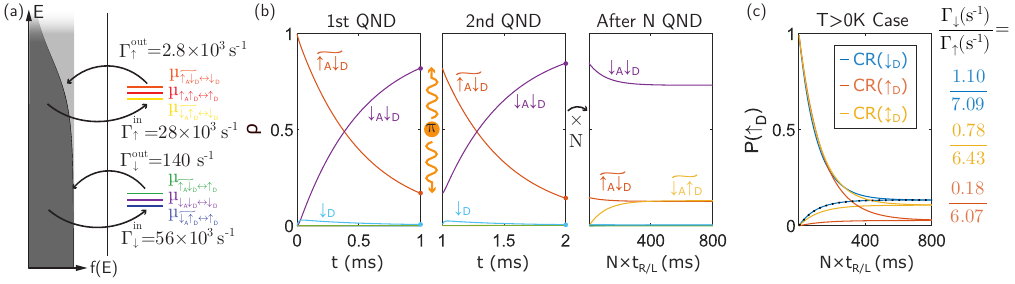}
    \caption{Example of error channels in the QND measurement of exchange-coupled spins with experimental parameters. (a) Chemical potentials with respect to a Fermi distribution in the lead for an off-centered read/load tuning representing the given set of experimentally-informed spin-dependent tunnel rates from \req{Eq:BaseRates}. For an electron in a magnetic field $B_0=\SI{1.77}{\tesla}$, where $\mu_\uparrow-\mu_\downarrow=2\mu_B\times\SI{1.77}{T}\approx\SI{50}{\giga\hertz}$, the ratios of spin-up and spin-down tunnel rates correspond to a temperature of roughly $\SI{600}{\milli\kelvin}$. (b) Simulations of the QND measurement with an initial $\downarrow_{\rm D}$, and driving a conditional rotation on the ancilla dependent on the data being $\downarrow_{\rm D}$ (CR$(\downarrow_{\rm D})$) for the tunnel rates in (a), and $s^2=\SI{2.5e-3}{}$. The first two panels show the state distribution as a function of time during the first two read/load periods. The last panel gives the final distribution after the Nth repetition, i.e. the last point of each QND step spaced by $t_{\rm R/L}=\SI{1}{\milli\second}$.  (c) Probability of finding the data in an $\uparrow_{\rm D}$ state (\req{Eq:Pup}) after repeated QND measurements as a function of time. The cases of driving the CR$(\uparrow_{\rm D})$, CR$(\downarrow_{\rm D})$, and alternating between the two (CR$(\updownarrow_{\rm D})$) are shown. The case of CR$(\downarrow_{\rm D})$ and an initial $\downarrow_{\rm D}$ discussed in (b) is highlighted by the black dotted line. The data qubit flipping rates on the right are extracted by fitting \req{Eq:Fitting}. The evolution of the data qubit is well described by these rates, independently of its initial state.}
    \label{Fig:3}
\end{figure*}

\begin{table}
\caption{ Data qubit flipping rates in the QND measurements using exchange-coupled donor electron spins additionally including a spin relaxation rate $\Gamma^{T_1}_{\uparrow}=\SI{1}{\per\second}$.}
\label{Tab:FlipRatesT1}
\begin{tabular}{c c c c}
 \hline
 & CR$(\downarrow_D)$ &  CR$(\uparrow_D)$ &  CR$(\updownarrow_D)$\\ 
 \hline\hline
$\Gamma_{\uparrow}$ (\SI{}{\per\second})& 7.11 & 6.47 & 6.70 \\ 
$\Gamma_{\downarrow}$ (\SI{}{\per\second})& 1.10 & 0.18 & 0.78 \\
\end{tabular}
\end{table}

For the analysis of donor electron spin experiments, we augment the model in the previous section by relaxing the assumption that the lead has a constant density of states \cite{morello2009architecture,mottonen2010probe}, and its chemical potential is perfectly centered between the $\mu_{\uparrow}$ and $\mu_{\downarrow}$ transition energies. \rfig{Fig:3} describes the repetitive quasi-QND readout protocol, using experimentally-informed values for the spin-dependent tunneling rates (\req{Eq:BaseRates}) given in \rfig{Fig:3}a. These rates are chosen to allow a direct comparison to experiments. The sketched off-centered tuning of the transition energies with respect to the Fermi distribution at the readout position can explain the ratio of tunnel in and out rates, while a non-constant density of states\cite{mottonen2010probe} $n(E)$ or tunnel coupling $t_0$ can motivate the differences between spin up and down rates. For an electron in a field of $B_0=\SI{1.77}{\tesla}$ as used in Ref.~\cite{pla2013high}, where $\mu_\uparrow-\mu_\downarrow=2\mu_B\times\SI{1.77}{T}\approx\SI{50}{\giga\hertz}$, the ratios of spin-up and spin-down tunnel rates correspond to a temperature of roughly $\SI{600}{\milli\kelvin}$. This value is seemingly high compared to the base temperature of dilution refrigerators, but consistent with effective electron temperatures observed in recent experiments\cite{johnson2022beating}.

\rfig{Fig:3}b shows the simulation of the QND protocol for this set of parameters, an initial $\downarrow_{\rm D}$, and driving the CR$(\downarrow_{\rm D})$, i.e. the same as \rfig{Fig:2}e but with the parameters in \rfig{Fig:3}a. We can see that now the initial peak belonging to the 1P state is much smaller, i.e. the system resides in a 1P state for a much shorter time, due to the fast tunnel-in rates. Additionally, the smaller ratio of $\Gamma^{\rm in}_{\downarrow}/\Gamma^{\rm in}_{\uparrow}=2$ decreases the probability to successfully reinitialize a $\downarrow_{\rm A}$ spin. For the chosen read/load window of $t_{\rm R/L}\SI{1}{\milli\second}$, this probability is 0.82. Looking at the state distribution after the Nth QND repetition, we can see that the probability of flipping the data from $\downarrow_{\rm D}$ to $\uparrow_{\rm D}$ is significantly increased compared to \rfig{Fig:2}e. This is due to the fact that the higher probability of accidentally loading an $\uparrow_{\rm A}$ spin increases the likelihood of the $\downarrow_{\rm D}\rightarrow\widetilde{\downarrow_{\rm A}\uparrow_{\rm D}}$ process. In the equilibrium state, the probability of finding the data in a $\downarrow_{\rm D}$ state (see \req{Eq:Pup}) is 0.13.

Plotting this quantity on a linear scale for all six considered cases, i.e. initial $\uparrow_{\rm D}$ or $\downarrow_{\rm D}$ with CR$(\uparrow_{\rm D})$, CR$(\downarrow_{\rm D})$, or CR$(\updownarrow_{\rm D})$, in \rfig{Fig:3}c, we recognize the same features as in \rfig{Fig:2}f. However now the equilibrium is shifted more towards $\uparrow_{\rm D}$ states. The data qubit flipping rates on right in \rfig{Fig:3}c are extracted by fitting \req{Eq:Fitting}. In comparison to the natural spin relaxation rate $\Gamma^{T_1}_{\uparrow}\sim\SI{1}{\per\second}$\cite{tenberg2019electron,hsueh2023hyperfine} of an donor-electron spin, the imperfections of the QND measurements are the dominant process. We verify that including the relaxation process in our simulations (see \req{Eq:LiouvillianeeT1}) has but the effect to slightly accelerate the $\Gamma_{\downarrow}$ rates. The precise results are given in \rtab{Tab:FlipRatesT1}.

\subsection{Nuclear spin hyperfine-coupled to a bound electron}

\begin{figure}
    \includegraphics[width=1.0\columnwidth]{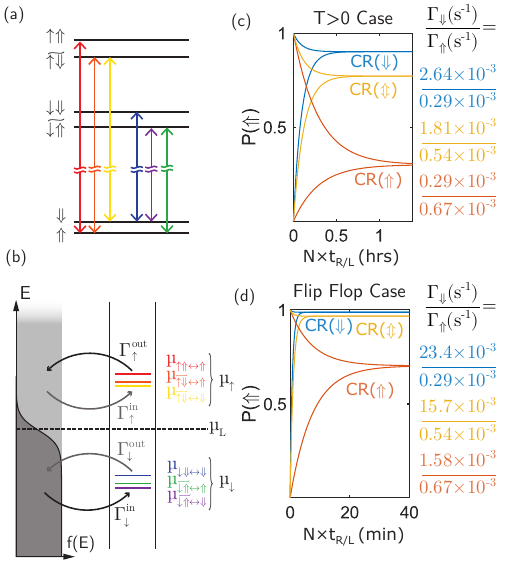}
    \caption{Error channels in the QND measurement of a nuclear spin via the hyperfine coupling to an electron, based on the example of a $^{31}$P atom with nuclear spin $I=1/2$. (a) The energy levels of \req{Eq:IsotropicHyperfine} are reordered compared to \rfig{Fig:2}a, since the nuclear spin energy splitting is much smaller than that of the electron: $\epsilon_{\rm e}\gg A_{\rm I}\gtrsim|\epsilon_{\rm n}|$, with $\epsilon_{\rm n}<0$. This gives a clear energetic distinction between allowed transitions (\req{Eq:SelectionRule}) of an $\uparrow$ or $\downarrow$ electron tunneling. (b) Transition energies with respect to an exemplary Fermi distribution in the lead for the read/load tuning. The resulting rates in \req{Eq:BaseRates} belonging to an $\uparrow$ or $\downarrow$ electron tunneling in or out are indicated. (c) Simulations of the QND measurement for $T>0$ (\req{Eq:Liouvillianen0K} and \req{Eq:LiouvillianenfK}) with the rates in \rfig{Fig:3}a, $s^2\approx10^{-6}$, and a read/load time of $t_{\rm R/L}=\SI{1}{\milli\second}$. Shown is the probability of finding the nucleus in a $\Uparrow$ state (\req{Eq:Pup}) after repeated QND measurements as a function of time. The cases of driving the CR$(\Downarrow)$, CR$(\Uparrow)$ alternating between the two CR$(\Updownarrow)$ are considered for an initial $\Uparrow$ and $\Downarrow$ state. The nuclear flipping rates on the right are extracted by fitting \req{Eq:Fitting}. (d) To reproduce the nuclear flipping rates in Ref. \cite{pla2013high} the $T>0$ model in (c) is augmented by including a flip-flop relaxation rate $\Gamma^{\rm ff}=\SI{53.3e-3}{\per\second}$ (\req{Eq:LiouvillianenT1}).}
    \label{Fig:4}
\end{figure}


The case of two exchange-coupled electrons considered so far, where the individual qubit energy splittings are only slightly different (the case $\epsilon_{\rm A}=\epsilon_{\rm D}$ is discussed in \rapp{App:C}), $\epsilon_{\rm A}\gtrsim\epsilon_{\rm D}$, makes the system prone to errors because (i) the small $\Delta\epsilon$ leads to a substantial hybridization of the $\uparrow_{\rm D}\downarrow_{\rm A}, \downarrow_{\rm D}\uparrow_{\rm A}$ states, and (ii) flip-flop transitions between data and ancilla qubits are almost energy-conserving. In this section, we discuss the case $\epsilon_{\rm A}\gg\epsilon_{\rm D}$, which is representative of a data qubit encoded in a donor nuclear spin\cite{pla2013high}, read out via an electron spin ancilla, coupled to the nucleus via Fermi contact hyperfine interaction. 

Mathematically, this system is similar to Heisenberg exchange-coupled spins (\req{Eq:Exchange}), since the isotropic Fermi contact hyperfine interaction takes the same Hamiltonian form:
\begin{align}
    \hat{H}=\epsilon_{\rm e}\hat{S}_{z}+\epsilon_{\rm n}\hat{I}_{z}+A_{\rm I}\Vec{S}\cdot\Vec{I},
    \label{Eq:IsotropicHyperfine}
\end{align}
where $A_{\rm I}$ is the hyperfine coupling strength and $\Vec{I}=(\hat{I}_{x},\hat{I}_{y},\hat{I}_{z})$ is the vector of nuclear spin operators in the basis $\{\ket{\Uparrow},\ket{\Downarrow}\}$ (we treat here the simple case where the donor is a $^{31}$P atom with nuclear spin $I=1/2$). The crucial difference is that the energy splitting of the ancilla electron is much larger than that of the data nucleus due to their vastly different gyromagnetic ratios\cite{pla2013high}, $|\gamma_{\rm e}/\gamma_{\rm n}B_0|>10^3$. With $\epsilon_{\rm e}\gg A_{\rm I}\gtrsim|\epsilon_{\rm n}|$ ($\epsilon_{\rm n}<0$ for a $^{31}$P atom), the $\uparrow$ states are energetically well separated from the $\downarrow$ ones in \rfig{Fig:4}a, while the nuclear splitting is two orders of magnitude smaller. As a result, the transition energies in \rfig{Fig:4}b are split depending on whether a spin up ($\mu_{\uparrow}$) or spin down ($\mu_{\downarrow}$) electron tunnels, independent of the nuclear spin state. This energy reordering now allows an initial $\Downarrow$ nucleus to be flipped in the zero-temperature loading process:
\begin{align}
    \begin{pmatrix}
    \dot{\rho}_{\widetilde{\downarrow\Uparrow}}\\
    \dot{\rho}_{\downarrow\Downarrow}\\
    \dot{\rho}_{\Downarrow}
    \end{pmatrix}
    =
    \begin{pmatrix}
    0&0&\Gamma^{\rm in}_{\downarrow}s^2\\
    0&0&\Gamma^{\rm in}_{\downarrow}\\
    0&0&-\Gamma^{\rm in}_{\downarrow}(1+s^2)
    \end{pmatrix}
    \begin{pmatrix}
    \rho_{\widetilde{\downarrow\Uparrow}}\\
    \rho_{\downarrow\Downarrow}\\
    \rho_{\Downarrow}
    \end{pmatrix}.
    \label{Eq:RateLoaden}
\end{align}
This is in contrast to \req{Eq:RateLoadee}, where only an $\uparrow_{\rm D}$ data could be flipped since the energy needed to excite an initially $\downarrow_{\rm D}$ data spin could not be provided at $\mu_{\downarrow}$. Using the parameters in the experiment of Ref.~\cite{pla2013high}, $A_{\rm I}=\SI{117.5}{\mega\hertz}$ and $\Delta\epsilon=(\gamma_{\rm e}-\gamma_{\rm n})B_0\approx\SI{50}{\giga\hertz}$ in a field of $B_0 = 1.77$~T, the resulting flip probability is
\begin{align}
    \frac{s^2}{1+s^2}\approx\left(\frac{A_{\rm I}}{2\Delta\epsilon}\right)^2=\SI{1.4e-6}{},
\end{align}
orders of magnitude smaller than in the exchange-coupled electron pair in \req{Eq:PJcase}. The nuclear spin readout is thus almost perfectly QND, even in presence of the noncommuting terms in \req{Eq:IsotropicHyperfine}. This is thanks to the hierarchy of energy scales $|\epsilon_{\rm e}|\gg |A|\gtrsim|\epsilon_{\rm n}|$, which ensures that the eigenstates of \req{Eq:IsotropicHyperfine} are almost exactly the product states ${\uparrow,\downarrow}\otimes{\Uparrow,\Downarrow}$. Furthermore, the isotropic nature of the interaction weakly hybridizes only the $\ket{\uparrow\Downarrow}, \ket{\downarrow\Uparrow}$ states, leaving $\ket{\uparrow\Uparrow}, \ket{\downarrow\Downarrow}$ as exact eigenstates unlike an anisotropic coupling discussed in the following section. Note that a positive $\epsilon_{\rm n}$ merely flips the order of the 1P states, while the above discussion remains valid.

\rfig{Fig:4}c shows simulations of the probabilities of finding the nucleus in the $\Uparrow$ state (compare \req{Eq:Pup}) after repeated QND measurements of the nucleus via the electron, using the combined Liouvillian of \req{Eq:Liouvillianen0K} and \req{Eq:LiouvillianenfK} with the rates from \rfig{Fig:3}a. The extracted flipping rates on the right are now on the order of $10^{-3}\,{\rm s}^{-1}$, i.e. on the order of a million QND measurements to flip the nucleus instead of a thousand for the exchange-coupled electron example. In general, we can see that the nucleus flips faster if the system is driven, i.e. if the electron frequently tunnels off and on. Consequently, driving the off-resonant transition will best preserve the nuclear spin state in a QND measurement. This is in contrast to the exchange-coupled spins, where driving the CR$(\downarrow_{\rm D})$ resulted in faster flipping rates for both spin configurations. As expected from \req{Eq:RateLoaden}, which describes the dominant error channel of a $\Downarrow$ nucleus being flipped as an electron is loaded, the fastest flipping rate is $\Gamma_\Downarrow$ in the case CR$(\Downarrow)$. In fact, in the zero-temperature limit, this would be the only mechanism causing nuclear spin flips.

Ref. \cite{pla2013high} performed nuclear QND measurements using the CR$(\Updownarrow)$ transitions, and found a rate $\Gamma_{\Uparrow}\approx\SI{0.6e-3}{\per\second}$, consistent with our prediction. However, they found a much faster $\Gamma_{\Uparrow}\approx \SI{15e-3}{\per\second}$, attributed to the electron-nuclear flip-flop relaxation process\cite{savytskyy2023electrically,hsueh2023hyperfine} $\widetilde{\uparrow\Downarrow}\rightarrow\widetilde{\downarrow\Uparrow}$. Including in our model for a flip-flop relaxation rate $\Gamma^{\rm ff} = \SI{53.3e-3}{\per\second}$ in \req{Eq:LiouvillianenfK} increases $\Gamma_{\Uparrow}$ to $\SI{15.73e-3}{\per\second}$, consistent with Ref. \cite{pla2013high}. The state evolutions of all cases (CR$(\Uparrow)$,CR$(\Downarrow)$,CR$(\Updownarrow)$) are shown in \rfig{Fig:4}d with the extracted nulcear flipping rates on the right. The CR$(\Downarrow)$ case constantly brings a $\Downarrow$ state to the $\widetilde{\uparrow\Downarrow}$ state prone to flip-flop relaxation, making $\Gamma_{\Downarrow}$ faster. The CR$(\Uparrow)$ case mostly avoids the $\widetilde{\uparrow\Downarrow}$ state and remains the preferred option. The intrinsic relaxation rate of a nuclear spin is immeasurably small\cite{savytskyy2023electrically} and needs not be included in the model.

Ref. \cite{pla2013high} also performed a resonant tunneling (RT) experiment described in detail \rapp{App:E}, where in each QND iteration $\mu_{\downarrow}$ is tuned in resonance with $\mu_{\rm L}$ for a \SI{0.7}{\milli\second} time period. During this time $\downarrow$ electrons repeatedly tunnel in and out at a high rate. The resulting nuclear flipping rates can be found in \rfig{Fig:A2}c. The CR$(\Updownarrow)$ case rates obtained with our model are in good agreement with the results in Ref. \cite{pla2013high}.

\subsection{Anisotropic hyperfine coupling}

\begin{figure}
    \includegraphics[width=1.0\columnwidth]{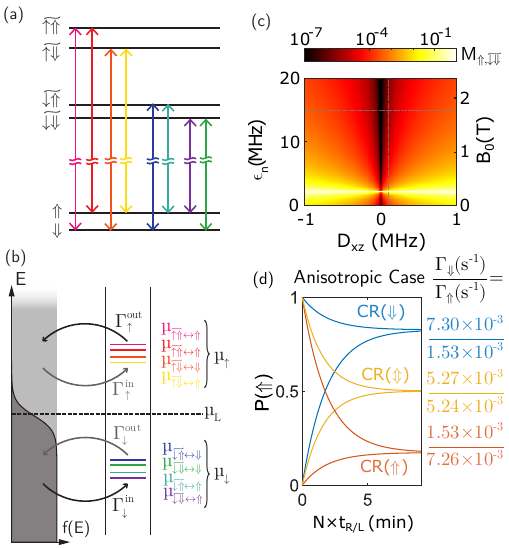}
    \caption{Error channels in the QND measurement of a nuclear spin in the presence of anisotropic hyperfine coupling with an electron. (a) Energy levels of \req{Eq:AnisotropicHyperfine} for  $\epsilon_{\rm e}\gg\epsilon_{\rm n}>A_{\rm I}>D_{ij}$. The anisotropic terms hybridize all spin states, such that \req{Eq:SelectionRule} now allows all possible transitions marked by the colored arrows between the loaded (top) and ionized (bottom) energy states. (b) Transition energies with respect to an exemplary Fermi distribution in the SET island for the read/load tuning. The resulting rates in \req{Eq:BaseRates} belonging to an $\uparrow$ or $\downarrow$ electron tunneling in or out are indicated. (c) $M_{\Uparrow,\widetilde{\downarrow\Downarrow}}$ (see \req{Eq:SelectionRule}) for eigenstates of \req{Eq:AnisotropicHyperfine} on a logarithmic scale as a function of $B_0$ and $D_{xz}$ with all other $D_{ij}=0$, $A_{\rm I}/2=\SI{2.254}{\mega\hertz}$, $\epsilon_{\rm n}=\SI{8.458}{\mega\hertz\per\tesla}\cdot B_0$, and $\epsilon_{\rm e}=\SI{27.97}{\giga\hertz\per\tesla}\cdot B_0$. The dashed lines mark $D_{xz}=\SI{106.2}{\kilo\hertz}$ and $B_0=\SI{1.77}{\tesla}$. (d) Simulations of QND measurements in the presence of anistropic hyperfine (not including $\Gamma^{\rm ff}$ or resonant tunneling) for the parameters above, the tunnel rates in \rfig{Fig:3}a, and a read/load time of $t_{\rm R/L}=\SI{1}{\milli\second}$. Shown are the probabilities of finding the nucleus in an $\Uparrow$ state (\req{Eq:Pup}) after repeated QND measurements as a function of time. The cases of driving the CR$(\Downarrow)$, CR$(\Uparrow)$ alternating between the two CR$(\Updownarrow)$ are considered for an initial $\Uparrow$ and $\Downarrow$ state. The nuclear flipping rates on the right are extracted by fitting \req{Eq:Fitting}.}
    \label{Fig:5}
\end{figure}


In the previous section, we considered a coupling that could hybridize the antiparallel spin states, i.e. $\ket{\uparrow\Downarrow}$ and $\ket{\downarrow\Uparrow}$, which introduced the error channels shown in \rfig{Fig:3}a. As these states are split by $\Delta\epsilon\approx\epsilon_{\rm e}$, which is much bigger than $A_{\rm I}$, this hybridization is weak and nuclear flipping rates slow.

In this section, we extend the discussion of nuclear readout by including an anisotropic hyperfine interaction\cite{slichter2013principles}. The electron-nuclear Hamiltonian becomes
\begin{align}
    \hat{H}=\epsilon_{\rm e}\hat{S}_{z}+\epsilon_{\rm n}\hat{I}_{z}+A_{\rm I}\Vec{S}\cdot\Vec{I}+\sum_{i,j}D_{ij}\hat{S}_i\hat{I}_j,
    \label{Eq:AnisotropicHyperfine}
\end{align}
where $D_{ij}$ describes the anisotropic dipolar coupling between electron and nucleus, and we sum over the Cartesian coordinates. In the case of donors in bulk silicon, $D_{ij}=0$ due to the spherical symmetry of the electron ground-state wavefunction. Breaking this symmetry, e.g. by an electric field or by strain, typically results in anistropic terms much smaller than the isotropic ones $A_{\rm I}\gg D_{ij}$\cite{wang2016characterizing,hile2018addressable}. This Hamiltonian describes many electron-nuclear spin systems, such as NV-diamond\cite{jiang2009repetitive,neumann2010single,robledo2011high,waldherr2011dark,dreau2013single,lovchinsky2016nuclear}, defects in SiC\cite{bourassa2020entanglement}, molecular magnets\cite{thiele2014electrically}, atoms on surfaces\cite{willke2018hyperfine}, rare-earth ions\cite{yang2022zeeman}, and $^{29}$Si nuclei coupled to donors\cite{ivey1975ground,pla2014coherent,mkadzik2020controllable} or quantum dots\cite{hensen2020silicon} in silicon. \req{Eq:AnisotropicHyperfine} hybridizes all spin states in \rfig{Fig:5}a, such that \req{Eq:SelectionRule} now allows all possible transitions between 1P and 2P states indicated by the arrows. As a result, the anisotropic hyperfine coupling term in \req{Eq:AnisotropicHyperfine} introduces new error channels in the QND measurement. 

Crucially, the $D_{xz}$ and $D_{yz}$ components hybridize the $\downarrow$ subspace, i.e. $\ket{\downarrow\Uparrow}$ and $\ket{\downarrow\Downarrow}$, as well as the $\uparrow$ one. The corresponding transitions in \rfig{Fig:5}b can flip the nuclear spin as a $\downarrow$ electron tunnels in, or an $\uparrow$ tunnels out, which routinely happens during the measurements. The amplitude of the transition matrix elements $M_{\rm 1P,2P}$ (\req{Eq:SelectionRule}) of this detrimental process depends on the ratio of diagonal vs off-diagonal elements in the respective subspace. Restricting ourselves to the $D_{xz}$ tensor component, \req{Eq:AnisotropicHyperfine} in the $\{\ket{\downarrow\Uparrow},\ket{\downarrow\Downarrow}\}$ subspace has the form
\begin{align}
H=\frac{1}{2}
\begin{pmatrix}
    -\epsilon_{\rm e}+(\epsilon_{\rm n}-A_{\rm I}/2)&D_{xz}/2\\
    D_{xz}/2&-\epsilon_{\rm e}-(\epsilon_{\rm n}-A_{\rm I}/2)
\end{pmatrix}.
\label{Eq:downSubspace}
\end{align}
We can interpret $D_{xz}$ as an effective field that tilts the quantization axis of the nuclear spin through the presence of the electron\cite{prokof2000theory}, effectively hybridizing the nuclear spin basis states. The smaller the nuclear energy splitting $\epsilon_{\rm n}-A_{\rm I}/2$, the stronger the hybridization. While $A_{\rm I}$ is typically only weakly tunable\cite{savytskyy2023electrically}, $\epsilon_{\rm n}=\gamma_{\rm n}B_0$ is a function of the applied magnetic field, which allows the degree of hybridization to be tuned in an experiment. \rfig{Fig:5}c shows calculations of $M_{\Uparrow,\widetilde{\downarrow\Downarrow}}$, i.e the $\downarrow$ subspace hybridization, as a function of $\epsilon_{\rm n}$ ($B$) and $D_{xz}$ on a logarithmic scale for the parameters of an exemplary  $^{29}$Si atom near a phosphorus donor in silicon\cite{ivey1975ground,pla2014coherent}: $A_{\rm I}/2=\SI{2.254}{\mega\hertz}$, $\gamma_{\rm n}=\SI{8.458}{\mega\hertz\per\tesla}$, and $\gamma_{\rm e}=\SI{27.97}{\giga\hertz\per\tesla}$. At the point where $\epsilon_{\rm n}=A_{\rm I}/2$ the hybridization is strong as expected. To reach the same degree of hybridization at a bigger splitting in \req{Eq:downSubspace} requires an increasingly stronger coupling $D_{xz}$. The $\uparrow$ subspace hybridization would be the strongest when $\epsilon_{\rm n}=-A_{\rm I}/2$. However, for a negative magnetic field everything is effectively flipped, in other words $\downarrow$ becomes $\uparrow$ and the hybridization in the $\uparrow$ subspace is always weaker for positive $\gamma_{\rm n}$. This model thus suggests measuring the nuclear flipping rate -- i.e. the deviation from QND measurement -- as a function of $B_0$ is an experimentally realizable way to quantify the anisotropy of the hyperfine interaction. This method would be particularly useful in cases where the anisotropy is very weak, and therefore would be difficult to observe directly in the resonance spectra, since the resonance frequency is only affected by off-diagonal elements to second order\cite{ivey1975ground}. 

For the chosen $^{29}$Si atom\cite{ivey1975ground} with $D_{xz}=\SI{106.2}{\kilo\hertz}$ and a field of $B=\SI{1.77}{\tesla}$ indicated in \rfig{Fig:5}c , where $M_{\Uparrow,\widetilde{\downarrow\Downarrow}}\approx4\times10^{-6}$ and the hybridization in the $\uparrow$ subspace $M_{\Uparrow,\widetilde{\uparrow\Downarrow}}\approx2\times10^{-6}$, we simulate the QND measurements using the Liouvillian in \rapp{App:E} and the rates in \rfig{Fig:3}a. \rfig{Fig:5}d shows the probability of finding the nucleus in the $\Uparrow$ state (compare \req{Eq:Pup}) after repeated QND measurements (not including $\Gamma^{\rm ff}$ or RT) with the extracted flipping rates on the right. Although the isotropic component of the hyperfine interaction $A_{\rm I}=\SI{4.5}{\mega\hertz}$ is much smaller than that of a  $^{31}$P atom ($A_{\rm I}=\SI{117.5}{\mega\hertz}$), the anisotropic components lead to faster flipping rates than in \rtab{Fig:4}c. This is a direct consequence of the fact that, for a $D_{xz}$-like coupling, the simple presence of the electron is sufficient to hybridize the nuclear spin states. In other words, unlike the case of an isotropic $A_{\rm I}$ coupling, here the hybridization does not require a process that includes an electron spin flip.

\section{Conclusion}
In this paper, we established a theoretical framework to understand and quantify the error channels in QND measurements in realistic spin systems. Errors are introduced when the interaction, needed to map the spin state of the data qubit onto the ancilla, entangles the two systems (exchange/hyperfine) or tilts the quantization axis of the data spin (anisotropic hyperfine). This enables transitions which flip the data spin as the ancilla tunnels off and on as part of the spin readout process. Their rates depend on the degree of hybridization and which transitions are energetically allowed. We analysed this dependency for three different cases.

In exchange-coupled spins, the fact that $\epsilon_{\rm A}\approx\epsilon_{\rm D}$ leads to a strong hybridization and fast error rates. The dominant error channel is the flip of an $\uparrow_{\rm D}$ as its energy can be transferred on to a $\downarrow_{\rm A}$ in the tunnel process. We showed that for a realistic set of parameters this process is an order of magnitude faster than the natural spin relaxation rate.

For a nuclear spin hyperfine-coupled to an electron, $\epsilon_{\rm e}\gg\epsilon_{\rm n}$ holds and the much weaker hybridization leads to lower error rates. The energy of the tunnel process now almost exclusively depends on the electron spin state. Using a realistic set of parameters for a $^{31}$P atom and including a flip-flop relaxation process, allowed us to reproduce the nuclear flipping rates measured in Ref.\cite{pla2013high}.

Finally, we showed that anisotropic hyperfine components can significantly hybridize the nuclear spin states and lead to faster flipping rates. As the degree of hybridization can be tuned by varying the nuclear splitting with a magnetic field, measuring the nuclear flipping rates as a function of field and its direction could be used to probe the anisotropy of the hyperfine coupling.

\section*{Acknowledgements}
The research was supported by the Australian Research Council (Grants No. CE170100012, DP210103769), the US Army Research Office (Contract no. W911NF-17-1-0200), and the Australian Department of Industry, Innovation and Science (Grant No. AUSMURI000002).  H.G.S. and I.F.d.F. Acknowledge support from the Sydney Quantum Academy.  The views and conclusions contained in this document are those of the authors and
should not be interpreted as representing the official policies, either expressed or implied, of the Army Research
Office or the U.S. Government. The U.S. Government is authorized to reproduce and distribute reprints for
Government purposes notwithstanding any copyright notation herein.
\FloatBarrier
\newpage

\appendix
\section{Explicit treatment of the lead}
\label{App:Lead}
We treat the lead as a continuum of states described by the Hamiltonian 
\begin{align}
    \hat{H}_{\rm L}=\sum_{k,\sigma}\epsilon_{k}\hat{c}^\dagger_{k\sigma}\hat{c}_{k\sigma},
    \label{Eq:Hlead}
\end{align}
where $\hat{c}_{k\sigma}^\dagger$ creates a particle with momentum $k$, spin $\sigma$, and energy $\epsilon_{k}$ on the lead. Then the tunneling between the lead and the ancilla is described by
\begin{align}
    \hat{H}_{\rm T}=\sum_{k,\sigma}t_0(\hat{a}_{\sigma}\hat{c}^\dagger_{k\sigma}+\hat{a}^\dagger_{\sigma}\hat{c}_{k\sigma}),
\end{align}
where we assume the tunnel coupling $t_{0}$ to be constant over the relevant energy range, i.e. the tunnel barrier is not changing much with energy (E).

We are now interested in processes, where a particle tunnels to (from) the lead causing transition between the one-particle (1P) and two-particle (2P) states of the data-ancilla system. Assuming that the lead is initially in the state $\ket{\rm L}$, the rate of the process $\ket{\rm L}\ket{\rm 1P}\rightarrow\hat{c}_{k\sigma}\ket{\rm L}\ket{\rm 2P}$, where a particle with energy $\epsilon_{k}$ and spin $\sigma$ tunnels to the ancilla, is given by
\begin{align}
        \Gamma^{k,\sigma}_{{\rm 1P}\rightarrow {\rm 2P}}=|\braket{{\rm L, 1P}|\hat{H}_{\rm T}\hat{c}_{k\sigma}|{\rm L, 2P}}|^2\delta(\epsilon_{k}-\mu_{{\rm 1P}\leftrightarrow {\rm 2P}})\\
        =|\braket{{\rm 1P}|t_0\hat{a}_{\sigma}|{2P}}|^2|\braket{{\rm L}|\hat{c}^\dagger_{k\sigma}\hat{c}_{k\sigma}|{\rm L}}|^2\delta(\epsilon_{k}-\mu_{{\rm 1P}\leftrightarrow {\rm 2P}}).
\end{align}
The last term describes the fact that the particle can only tunnel if it carries the energy equal to the chemical potential of the transition, i.e. $\epsilon_{k}=\mu_{{\rm 2P}\leftrightarrow {\rm 1P}}$. The term $|\braket{{\rm L}|\hat{c}^\dagger_{k\sigma}\hat{c}_{k\sigma}|{\rm L}}|^2$ is 1 if the state with energy $\epsilon_{k}$ and spin $\sigma$ was initially occupied and 0 otherwise. By taking the thermal average over all states $\ket{\rm L}$, we replace this term by the Fermi distribution $f(\epsilon_{k})$, i.e. the fraction of states with energy $\epsilon_{k}$ that is on average occupied at a given temperature. We may do so as the energy $\epsilon_{k}$ does not depend on the spin. Summing over all spins $\sigma$ and momenta $k$ we get
\begin{align}
        &\Gamma_{{\rm 1P}\rightarrow {\rm 2P}}=\sum_{k,\sigma}\Gamma^{k,\sigma}_{{\rm 1P}\rightarrow {\rm 2P}}\\
        &=\sum_{k,\sigma}|\braket{{\rm 1P}|t_0\hat{a}_{\sigma}|{2P}}|^2f(\epsilon_{k})\delta(\epsilon_{k}-\mu_{{\rm 1P}\leftrightarrow {\rm 2P}})\\
        &=|\braket{{\rm 1P}|\sum_{\sigma}t_0\hat{a}_{\sigma}|{2P}}|^2f(\mu_{{\rm 1P}\leftrightarrow {\rm 2P}})n(\mu_{{\rm 1P}\leftrightarrow {\rm 2P}}),
\end{align}
where $n(E)$ is the density of states counting the number of states in the lead at energy $E$. We have arrived at \req{Eq:LoadRate} of the main text.

To derive \req{Eq:UnloadRate}, we start by looking at the tunnel process $\ket{\rm L}\ket{\rm 2P}\rightarrow\hat{c}^\dagger_{k\sigma}\ket{\rm L}\ket{\rm 1P}$ with rate
\begin{align}
        &\Gamma^{k,\sigma}_{{\rm 2P}\rightarrow {\rm 1P}}=|\braket{{\rm L, 2P}|\hat{H}_{\rm T}\hat{c}^\dagger_{k\sigma}|{\rm L, 1P}}|^2\delta(\epsilon_{k}-\mu_{{\rm 1P}\leftrightarrow {\rm 2P}})\\
        &=|\braket{{\rm 2P}|t_0\hat{a}^\dagger_{\sigma}|{1P}}|^2|\braket{{\rm L}|\hat{c}_{k\sigma}\hat{c}^\dagger_{k\sigma}|{\rm L}}|^2\delta(\epsilon_{k}-\mu_{{\rm 1P}\leftrightarrow {\rm 2P}})\\
        &=|\braket{{\rm 2P}|t_0\hat{a}^\dagger_{\sigma}|{1P}}|^2|\braket{{\rm L}|1-\hat{c}^\dagger_{k\sigma}\hat{c}_{k\sigma}|{\rm L}}|^2\delta(\epsilon_{k}-\mu_{{\rm 1P}\leftrightarrow {\rm 2P}}).
\end{align}
From here we can follow the same steps as above to arrive at \req{Eq:UnloadRate}.
\section{Liouvillian of Heisenberg exchange coupled spins}
\label{App:A}
In the basis $\{\uparrow_{\rm A}\uparrow_{\rm D},\widetilde{\uparrow_{\rm A}\downarrow_{\rm D}},\widetilde{\downarrow_{\rm A}\uparrow_{\rm D}},\downarrow_{\rm A}\downarrow_{\rm D},\uparrow_{\rm D},\downarrow_{\rm D}\},$ the Liouvillian of two Heisenberg exchange-coupled spins in the zero-temperature limit is given by
\begin{align}
\mathcal{L}_{T=0}=
 \begin{pmatrix}
-\Gamma^{\rm out}_{\uparrow}&0&0&0&0&0\\
0&-\Gamma^{\rm out}_{\uparrow}c^2&0&0&\Gamma^{\rm in}_{\downarrow}s^2&0\\
0&0&-\Gamma^{\rm out}_{\uparrow}s^2&0&\Gamma^{\rm in}_{\downarrow}c^2&0\\
0&0&0&0&0&\Gamma^{\rm in}_{\downarrow}\\
\Gamma^{\rm out}_{\uparrow}&0&0&0&-\Gamma^{\rm in}_{\downarrow}&0\\
0&\Gamma^{\rm out}_{\uparrow}c^2&\Gamma^{\rm out}_{\uparrow}s^2&0&0&-\Gamma^{\rm in}_{\downarrow}
\end{pmatrix}.
\label{Eq:Liouvillian2e0K}
\end{align}
Here, the second-last column describes the loading process in \req{Eq:RateLoadee}. In the $T>\SI{0}{\kelvin}$ case, we get the following additional matrix elements
\begin{align}
\begin{split}
\mathcal{L}_{T>0}=
\begin{pmatrix}
0&0&0&0&\Gamma^{\rm in}_{\uparrow}&0\\
0&-\Gamma^{\rm out}_{\downarrow}s^2&0&0&0&\Gamma^{\rm in}_{\uparrow}c^2\\
0&0&-\Gamma^{\rm out}_{\downarrow}c^2&0&&\Gamma^{\rm in}_{\uparrow}s^2\\
0&0&0&-\Gamma^{\rm out}_{\downarrow}&0&0\\
0&\Gamma^{\rm out}_{\downarrow}s^2&\Gamma^{\rm out}_{\downarrow}c^2&0&-\Gamma^{\rm in}_{\uparrow}&0\\
0&0&0&\Gamma^{\rm out}_{\downarrow}&0&-\Gamma^{\rm in}_{\uparrow}
\end{pmatrix}.
\end{split}
\label{Eq:Liouvillian2efK}
\end{align}
The $T_1$ spin relaxation process with rate $\Gamma^{T_1}_{\uparrow}$ is described by
\begin{align}
\mathcal{L}_{T_1}=
\begin{pmatrix}
-2\Gamma^{T_1}_{\uparrow}&0&0&0&0&0\\
\Gamma^{T_1}_{\uparrow}&-\Gamma^{T_1}_{\uparrow}&0&0&0&0\\
\Gamma^{T_1}_{\uparrow}&0&-\Gamma^{T_1}_{\uparrow}&0&0&0\\
0&\Gamma^{T_1}_{\uparrow}&\Gamma^{T_1}_{\uparrow}&0&0&0\\
0&0&0&0&-\Gamma^{T_1}_{\uparrow}&0\\
0&0&0&0&\Gamma^{T_1}_{\uparrow}&0
\end{pmatrix}.
\label{Eq:LiouvillianeeT1}
\end{align}

\section{Flipping rates and equilibrium state}
\label{App:B}
Given that an $\uparrow$ is flipped with rate $\Gamma_{\uparrow}$ and a $\downarrow$ with $\Gamma_{\downarrow}$, their time evolution is described by the following rate equation
\begin{align}
    \begin{pmatrix}
    \dot{\rho}_{\uparrow}\\
    \dot{\rho}_{\downarrow}
    \end{pmatrix}
    =
    \begin{pmatrix}
    -\Gamma_{\uparrow}&\Gamma_{\downarrow}\\
    \Gamma_{\uparrow}&-\Gamma_{\downarrow}\\
    \end{pmatrix}
    \begin{pmatrix}
    \rho_{\uparrow}\\
    \rho_{\downarrow}
    \end{pmatrix}.
    \label{Eq:Rate}
\end{align}
For an initial $\rho_{\uparrow}(0)=1$ we get the solution
\begin{align}
    \rho_{\uparrow}(t)=&\frac{\Gamma_{\downarrow}}{\Gamma_{\uparrow}+\Gamma_{\downarrow}}\left(1+\frac{\Gamma_{\uparrow}}{\Gamma_{\downarrow}}e^{-(\Gamma_{\uparrow}+\Gamma_{\downarrow})t}\right),\\
    \rho_{\downarrow}(t)=&\frac{\Gamma_{\uparrow}}{\Gamma_{\uparrow}+\Gamma_{\downarrow}}\left(1-e^{-(\Gamma_{\uparrow}+\Gamma_{\downarrow})t}\right),
    \label{Eq:Fitting}
\end{align}
which reaches the equilibrium state
\begin{align}
    &\rho_{\uparrow}(t=\infty)=\frac{\Gamma_{\downarrow}}{\Gamma_{\uparrow}+\Gamma_{\downarrow}},\\
    &\rho_{\downarrow}(t=\infty)=\frac{\Gamma_{\uparrow}}{\Gamma_{\uparrow}+\Gamma_{\downarrow}}.
\end{align}
This means that the ratio of state occupation in equilibrium equals the ratios of the flipping rates.

\section{Exchange-coupled spins with $\epsilon_{\rm A}=\epsilon_{\rm D}$}
\label{App:C}
\begin{figure}[t]
    \includegraphics[width=1.0\columnwidth]{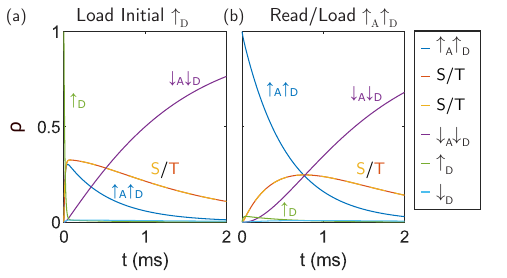}
    \caption{Simulations of a load/read period in the $\epsilon_{\rm A}=\epsilon_{\rm D}$ case, i.e. $s^2=c^2=1/2$, and the tunnel rates in \rfig{Fig:3}a. Shown is the state distribution as a function of time. In (a) the initial state is $\uparrow_{\rm D}$, while in (b) it is $\uparrow_{\rm A}\uparrow_{\rm D}$.}
    \label{Fig:A1}
\end{figure}

In the case where the two coupled electrons have the same individual energy splitting, $\epsilon_{\rm A}=\epsilon_{\rm D}$, the hybridized states in     \req{Eq:HybridStatedu} and \req{Eq:HybridStateud} become
\begin{align}
    \ket{S}&=\frac{1}{\sqrt{2}}(\ket{\uparrow_{\rm A}\downarrow_{\rm D}}-\ket{\downarrow_{\rm A}\uparrow_{\rm D}})
    \label{Eq:HybridStateS}
    \\
    \ket{T}&=\frac{1}{\sqrt{2}}(\ket{\downarrow_{\rm A}\uparrow_{\rm D}}+\ket{\uparrow_{\rm A}\downarrow_{\rm D}}).
    \label{Eq:HybridStateT}
\end{align}
This makes it unreasonable to attempt the QND measurement, since an $\uparrow_{\rm D}$ can be flipped within a single load cycle as shown in \rfig{Fig:A1}a. Here we use the tunnel rates in \rfig{Fig:3}a with $s=c=1/\sqrt{2}$. The initial $\uparrow_{\rm D}$ is quickly loaded into either a $S$ or $T$ state. However, these states still contain a $\mu_{\uparrow}$ energy quantum, which can be released by one of the electrons tunneling out, leaving a flipped $\downarrow_{\rm D}$ behind. Finally, the ancilla is reloaded into $\downarrow_{\rm A}\downarrow_{\rm D}$. Note that the simulated load time was doubled to compensate for the $S$ and $T$ tunnel times that are twice as long for $c^2=s^2=1/2$.

This process could also explain the experimentally observed increase of the visibility of the $\uparrow_{\rm A}\uparrow_{\rm D}$ state in exchange-coupled phosphorus donors\cite{madzik2021conditional} when the nuclear spins are parallel, i.e. $\epsilon_{\rm A}=\epsilon_{\rm D}$. During the read/load simulated in \rfig{Fig:A1}b the ancilla will tunnel off and back on twice, first while $\uparrow_{\rm A}\uparrow_{\rm D}\rightarrow\uparrow_{\rm D}\rightarrow S/T$ and secondly while $S/T\rightarrow\downarrow_{\rm D}\rightarrow\downarrow_{\rm A}\downarrow_{\rm D}$. This gives the charge sensor two chances to detect a signal instead of just once, which can make a visible difference when the single-shot readout contrast is not perfect.

\section{Liouvillian of a nucleus hyperfine-coupled to an electron}
\label{App:D}

In the basis $\{\uparrow\Uparrow,\widetilde{\uparrow\Downarrow},\widetilde{\downarrow\Uparrow},\downarrow\Downarrow,\Uparrow,\Downarrow\}$, the $T=0$ Liouvillian of a nucleus hyperfine-coupled to an electron is
\begin{align}
\mathcal{L}_{T=0}=
\begin{pmatrix}
-\Gamma^{\rm out}_{\uparrow}&0&0&0&0&0\\
0&-\Gamma^{\rm out}_{\uparrow}&0&0&0&0\\
0&0&0&0&\Gamma^{\rm in}_{\downarrow}c^2&\Gamma^{\rm in}_{\downarrow}s^2\\
0&0&0&0&0&\Gamma^{\rm in}_{\downarrow}\\
\Gamma^{\rm out}_{\uparrow}&\Gamma^{\rm out}_{\uparrow}s^2&0&0&-\Gamma^{\rm in}_{\downarrow}c^2&0\\
0&\Gamma^{\rm out}_{\uparrow}c^2&0&0&0&-\Gamma^{\rm in}_{\downarrow}(1+s^2)
\end{pmatrix}.
\label{Eq:Liouvillianen0K}
\end{align}
In the $T>\SI{0}{\kelvin}$ case, we get the following additional matrix elements
\begin{align}
\mathcal{L}_{T>0}=
\begin{pmatrix}
0&0&0&0&\Gamma^{\rm in}_{\uparrow}&0\\
&0&0&0&\Gamma^{\rm in}_{\uparrow}s^2&\Gamma^{\rm in}_{\uparrow}c^2\\
0&0&-\Gamma^{\rm out}_{\downarrow}&0&0&0\\
0&0&0&-\Gamma^{\rm out}_{\downarrow}&0&0\\
0&0&\Gamma^{\rm out}_{\downarrow}c^2&0&-\Gamma^{\rm in}_{\uparrow}(1+s^2)&0\\
0&0&\Gamma^{\rm out}_{\downarrow}s^2&\Gamma^{\rm out}_{\downarrow}&0&-\Gamma^{\rm in}_{\uparrow}c^2
\end{pmatrix}.
\label{Eq:LiouvillianenfK}
\end{align}
Relaxation processes are described by
\begin{align}
\mathcal{L}_{T_1}=
\begin{pmatrix}
-\Gamma^{T_1}_{\uparrow}&0&0&0&0&0\\
\Gamma^{T_1}_{\uparrow}s^2&-\Gamma^{T_1}_{\uparrow}c^2-\Gamma^{\rm ff}&0&0&0&0\\
\Gamma^{T_1}_{\uparrow}c^2&\Gamma^{\rm ff}&-\Gamma^{T_1}_{\uparrow}s^2&0&0&0\\
0&\Gamma^{T_1}_{\uparrow}c^2&\Gamma^{T_1}_{\uparrow}s^2&0&0&0\\
0&0&0&0&0&0\\
0&0&0&0&0&0
\end{pmatrix}.
\label{Eq:LiouvillianenT1}
\end{align}
where we also included the flip-flop relaxation process\cite{pla2013high,savytskyy2023electrically} with rate $\Gamma^{\rm ff}$.

\section{Nuclear flipping via resonant electron tunneling}
\label{App:E}

\begin{figure}
    \includegraphics[width=1.0\columnwidth]{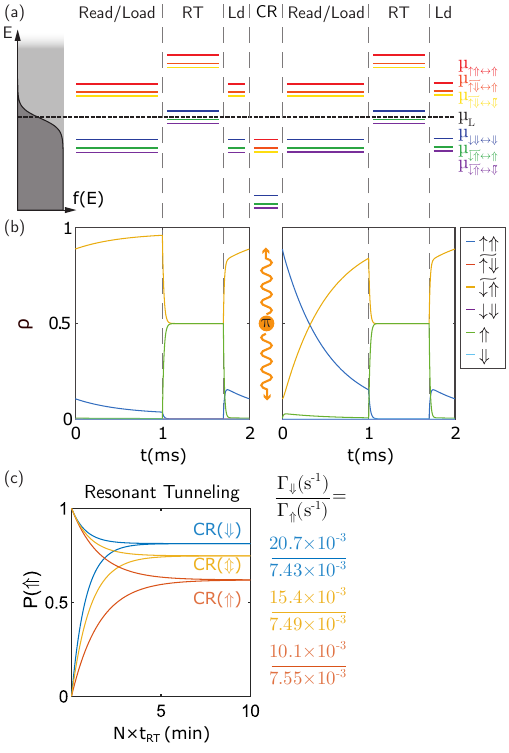}
    \caption{Nuclear spin flipping during resonant electron tunneling in a donor system. (a) Illustration of two consecutive resonant tunneling (RT) experiment iterations. Left: tuning of the transition energies with respect to an exemplary Fermi distribution of the SET island during the different stages of the RT iterations. Driving pluses are followed by a \SI{1}{\milli\second} read/load window, a $\SI{0.7}{\milli\second}$ RT stage, and a $\SI{0.3}{\milli\second}$ load time to reinitialize a $\downarrow$ electron. A RT iteration has a total length of $t_{\rm RT}=\SI{2}{\milli\second}$. During read and load periods we have the rates in \rfig{Fig:3}a, while during RT periods, where $\mu_{\downarrow}$ is tuned in resonance with $\mu_{\rm L}$, $\Gamma^{\rm in}_{\downarrow}=\Gamma^{\rm out}_{\downarrow}=\SI{28e3}{\per\second}$, $\Gamma^{\rm in}_{\uparrow}=\SI{140}{\per\second}$, and $\Gamma^{\rm out}_{\uparrow}=\SI{56e3}{\per\second}$. $s^2\approx10^{-6}$ and $\Gamma^{\rm ff}=\SI{53.3e-3}{\per\second}$. (b) Resulting state distribution as a function of time. In the RT period an equilibrium of an equally likely neutral (electron loaded) or ionized (electron unloaded) donor is quickly reached. (c) Probabilities of finding the nucleus in a $\Uparrow$ state (\req{Eq:Pup}) after repeated RT iterations as a function of time. The cases of driving the CR$(\Downarrow)$, CR$(\Uparrow)$ alternating between the two CR$(\Updownarrow)$ are considered for an initial $\Uparrow$ and $\Downarrow$ state. The nuclear flipping rates on the right are extracted fitting \req{Eq:Fitting}.}
    \label{Fig:A2}
\end{figure}

Ref. \cite{pla2013high} performed a QND measurement on a $^{31}$P nuclear spin hyperfine-coupled to an electron, while including a resonant tunneling (RT) window for the electron, whereby the $\downarrow$ electron state was tuned in resonance with the electrochemical potential of the lead,  $\mu_{\downarrow}=\mu_{\rm L}$ (see RT in \rfig{Fig:A2}a). This tuning yields $\Gamma^{\rm out}_{\uparrow}>\Gamma^{\rm out}_{\downarrow}=\Gamma^{\rm in}_{\downarrow}\gg\Gamma^{\rm in}_{\uparrow}$, and results in $\downarrow$ electrons randomly and frequently tunneling on and off the donor (akin to random telegraph signals\cite{house2013detection}). The process is described by the following rate equation
\begin{align}
    \begin{pmatrix}
    \dot{\rho}_{\widetilde{\downarrow\Uparrow}}\\
    \dot{\rho}_{\downarrow\Downarrow}\\
    \dot{\rho}_{\Uparrow}\\
    \dot{\rho}_{\Downarrow}
    \end{pmatrix}
    =
    \begin{pmatrix}
    -\Gamma^{\rm t}_{\downarrow}&0&\Gamma^{\rm t}_{\downarrow}c^2&\Gamma^{\rm rt}_{\downarrow}s^2\\
    0&-\Gamma^{\rm rt}_{\downarrow}&0&\Gamma^{\rm rt}_{\downarrow}\\
    \Gamma^{\rm rt}_{\downarrow}c^2&0&-\Gamma^{\rm rt}_{\downarrow}c^2&0\\
    \Gamma^{\rm rt}_{\downarrow}s^2&\Gamma^{\rm rt}_{\downarrow}&0&-\Gamma^{\rm rt}_{\downarrow}(1+s^2)\\
    \end{pmatrix}
    \begin{pmatrix}
    \rho_{\widetilde{\downarrow\Uparrow}}\\
    \rho_{\downarrow\Downarrow}\\
    \rho_{\Uparrow}\\
    \rho_{\Downarrow}
    \end{pmatrix},
    \label{Eq:RateRT}
\end{align}
where $\Gamma^{\rm rt}_{\downarrow}=\Gamma^{\rm in/out}_{\downarrow}$. Due to $\Gamma^{\rm out}_{\downarrow}=\Gamma^{\rm in}_{\downarrow}$, the system will quickly reach an equilibrium $\rho_{\Uparrow}=\rho_{\widetilde{\downarrow\Uparrow}}$ and $\rho_{\Downarrow}=\rho_{\downarrow\Downarrow}$, which allows us to simplify \req{Eq:RateRT} to
\begin{align}
    \begin{pmatrix}
    \dot{\rho}_{\Uparrow}\\
    \dot{\rho}_{\Downarrow}
    \end{pmatrix}
    =
    \begin{pmatrix}
    -\Gamma^{\rm rt}_{\downarrow}s^2&\Gamma^{\rm rt}_{\downarrow}s^2\\
    \Gamma^{\rm rt}_{\downarrow}s^2&-\Gamma^{\rm rt}_{\downarrow}s^2\\
    \end{pmatrix}
    \begin{pmatrix}
    \rho_{\Uparrow}\\
    \rho_{\Downarrow}
    \end{pmatrix}.
    \label{Eq:RateRTSimple}
\end{align}
\req{Eq:RateRTSimple} means that during the RT period both nuclear spin configurations will be flipped at a rate $\Gamma^{\rm rt}_{\downarrow}s^2$, i.e. a significantly faster flipping of the $\Uparrow$ nucleus. 

\rfig{Fig:A2}a illustrates the tuning of the electrochemical potentials during a repetition of the RT experiment\cite{pla2013high}. Driving pluses are followed by a \SI{1}{\milli\second} read/load window as in the usual QND experiment. Here, this is followed by a $\SI{0.7}{\milli\second}$ RT window and a $\SI{0.3}{\milli\second}$ load time to reinitialize a $\downarrow$ electron.

\rfig{Fig:A2}b shows simulations for pulses alternating between case 1 and 2, an initial $\Uparrow$ nucleus. In the RT window we assume $\Gamma^{\rm rt}_{\downarrow}=\SI{28e3}{\per\second}$, $\Gamma^{\rm in}_{\uparrow}=\SI{140}{\per\second}$, and $\Gamma^{\rm out}_{\uparrow}=\SI{56e3}{\per\second}$. Before the first shown read/load window, a  CR$(\Downarrow)$ leaves the $\widetilde{\downarrow\Uparrow}$ initialized in a previous iteration unchanged. Reading a $\downarrow$ electron only improves the state preparation, i.e. decreases the remaining $\uparrow$ probability, through a longer wait time. At the following RT tuning, we quickly reach the   $\rho_{\Uparrow}=\rho_{\widetilde{\downarrow\Uparrow}}=0.5$ equilibrium discussed above at which the nucleus is flipped at the rate $\Gamma_{\downarrow}^{\rm rt}s^2$. Before applying a driving pulse, now conditional on a $\Uparrow$ nucleus (CR$(\Uparrow)$), a short loading window initializes a $\downarrow$ electron state. In the subsequent read/load window, the $\uparrow$ electron tunnels out and the donor is quickly reloaded with a $\downarrow$ followed by the same RT and reload processes. The probabilities of finding the nucleus in an $\Uparrow$ after many repetitions of this sequence for the cases of driving the CR$(\Downarrow)$, CR$(\Uparrow)$ or alternating between the two CR$(\Updownarrow)$, for an initial $\Uparrow$ and $\Downarrow$ state, are shown in \rfig{Fig:A2}c with the extracted flipping rates on the right.

\section{Liouvillian for anisotropic hyperfine coupling}
\label{App:F}
In the basis of the eigenstates of \req{Eq:AnisotropicHyperfine} $\{\widetilde{\uparrow\Uparrow},\widetilde{\uparrow\Downarrow},\widetilde{\downarrow\Uparrow},\widetilde{\downarrow\Downarrow},\Uparrow,\Downarrow\}$ the $T=0$ Liouvillian of nucleus coupled to an electron via anisotropic hyperfine is
\begin{widetext}
\begin{align}
\mathcal{L}_{T=0}=
\begin{pmatrix}
-\Gamma^{\rm out}_{\widetilde{\uparrow\Uparrow}\rightarrow\Uparrow}-\Gamma^{\rm out}_{\widetilde{\uparrow\Uparrow}\rightarrow\Uparrow}&0&0&0&0&0\\
0&-\Gamma^{\rm out}_{\widetilde{\uparrow\Downarrow}\rightarrow\Uparrow}-\Gamma^{\rm out}_{\widetilde{\uparrow\Downarrow}\rightarrow\Uparrow}&0&0&0&0\\
0&0&0&0&\Gamma^{\rm in}_{\Uparrow\rightarrow\widetilde{\downarrow\Uparrow}}&\Gamma^{\rm in}_{\Downarrow\rightarrow\widetilde{\downarrow\Uparrow}}\\
0&0&0&0&\Gamma^{\rm in}_{\Uparrow\rightarrow\widetilde{\downarrow\Downarrow}}&\Gamma^{\rm in}_{\Downarrow\rightarrow\widetilde{\downarrow\Downarrow}}\\
\Gamma^{\rm out}_{\widetilde{\uparrow\Uparrow}\rightarrow\Uparrow}&\Gamma^{\rm out}_{\widetilde{\uparrow\Downarrow}\rightarrow\Uparrow}&0&0&-\Gamma^{\rm in}_{\Uparrow\rightarrow\widetilde{\downarrow\Uparrow}}-\Gamma^{\rm in}_{\Uparrow\rightarrow\widetilde{\downarrow\Downarrow}}&0\\
\Gamma^{\rm out}_{\widetilde{\uparrow\Uparrow}\rightarrow\Downarrow}&\Gamma^{\rm out}_{\widetilde{\uparrow\Downarrow}\rightarrow\Downarrow}&0&0&0&-\Gamma^{\rm in}_{\Downarrow\rightarrow\widetilde{\downarrow\Uparrow}}-\Gamma^{\rm in}_{\Downarrow\rightarrow\widetilde{\downarrow\Downarrow}}
\end{pmatrix},
\label{Eq:Liouvilliananiso0K}
\end{align}
\end{widetext}
where the rates are given by
\begin{subequations}
\begin{align}
    &\Gamma^{\rm out}_{\rm 2P\rightarrow1P}=\Gamma^{\rm out}_{\uparrow} M_{\rm 1P,2P},\\
    &\Gamma^{\rm in}_{\rm 1P\rightarrow2P}=\Gamma^{\rm in}_{\downarrow}M_{\rm 1P,2P}
\end{align}
\end{subequations}
In the $T>\SI{0}{\kelvin}$ case, we get the following additional matrix elements
\begin{widetext}
\begin{align}
\mathcal{L}_{T>0}=
\begin{pmatrix}
0&0&0&0&\Gamma^{\rm in}_{\Uparrow\rightarrow\widetilde{\uparrow\Uparrow}}&\Gamma^{\rm in}_{\Downarrow\rightarrow\widetilde{\uparrow\Uparrow}}\\
0&0&0&0&\Gamma^{\rm in}_{\Uparrow\rightarrow\widetilde{\uparrow\Downarrow}}&\Gamma^{\rm in}_{\Downarrow\rightarrow\widetilde{\uparrow\Downarrow}}\\
0&0&-\Gamma^{\rm out}_{\widetilde{\downarrow\Uparrow}\rightarrow\Uparrow}-\Gamma^{\rm out}_{\widetilde{\downarrow\Uparrow}\rightarrow\Downarrow}&0&0&0\\
0&0&0&-\Gamma^{\rm out}_{\widetilde{\downarrow\Downarrow}\rightarrow\Uparrow}-\Gamma^{\rm out}_{\widetilde{\downarrow\Downarrow}\rightarrow\Downarrow}&0&0\\
0&0&\Gamma^{\rm out}_{\widetilde{\downarrow\Uparrow}\rightarrow\Uparrow}&\Gamma^{\rm out}_{\widetilde{\downarrow\Downarrow}\rightarrow\Uparrow}&-\Gamma^{\rm in}_{\Uparrow\rightarrow\widetilde{\uparrow\Uparrow}}-\Gamma^{\rm in}_{\Uparrow\rightarrow\widetilde{\uparrow\Downarrow}}&0\\
0&0&\Gamma^{\rm out}_{\widetilde{\downarrow\Uparrow}\rightarrow\Downarrow}&\Gamma^{\rm out}_{\widetilde{\downarrow\Downarrow}\rightarrow\Downarrow}&0&-\Gamma^{\rm in}_{\Downarrow\rightarrow\widetilde{\uparrow\Uparrow}}-\Gamma^{\rm in}_{\Downarrow\rightarrow\widetilde{\uparrow\Downarrow}}
\end{pmatrix},
\label{Eq:LiouvilliananisofK}
\end{align}
\end{widetext}
where the rates are given by
\begin{subequations}
\begin{align}
    &\Gamma^{\rm in}_{\rm 1P\rightarrow2P}=\Gamma^{\rm in}_{\uparrow}M_{\rm 1P,2P},\\
    &\Gamma^{\rm out}_{\rm 2P\rightarrow1P}=\Gamma^{\rm out}_{\downarrow}M_{\rm 1P,2P}
\end{align}
\end{subequations}

\bibliography{citations} 

\end{document}